\documentclass[a4paper,11pt]{article}

\usepackage{jheppub} 
\newcommand{\beq}{\begin{equation}}
\newcommand{\eeq}{\end{equation}}
\newcommand\bO{\bar \CO} \newcommand\C{U_F} \newcommand\U{{\rm U}}
\newcommand\CV{{\cal V}} \newcommand\CN{{\cal N}}

 \newcommand\re[1]{({\ref{#1}})} \def\be{\begin{eqnarray} }
 \def\ee{\end{eqnarray}} \def\no{\nonumber} \def\la{\label}

 \def\Tr{{\rm Tr}}

\def\d{\delta}
\def\ket{ | 0 \rangle} 
 \def\({\left(} \def\){\right)} \def\<{\langle} \def\>{\rangle}
 \def\[{\left[} \def\]{\right]}  \def\hf{
 {\textstyle{1\over 2}} } 
 \def\CA{{\mathcal A}}   \def\CO{{
 \mathcal{ O} }}  
 \def\CH{ {\cal H}} \def\a{\alpha} \def\b{\beta} \def\g{\gamma}
 \def\e{\epsilon} \def\uu{ { {\{ u\}} }} \def\vv{ { \{ v\}}}
\def\U{ {\rm U}}
 
 \def\R{{\cal R}}
 \def\l{\lambda}

\title{ String Bits and the Spin Vertex}

\author{Yunfeng Jiang,}
\author{Ivan Kostov,}
\author{Andrei Petrovskii,}
\author{Didina Serban}

\emailAdd{yunfeng.jiang,ivan.kostov,andrei.petrovskii,didina.serban@cea.fr}

\affiliation{Institut de Physique Th\'eorique, DSM, CEA, URA2306
CNRS\\Saclay, F-91191 Gif-sur-Yvette, France}

\abstract{We initiate a novel formalism for computing correlation
functions of trace operators in the planar $\mathcal{N}=4$ SYM theory.
The central object in our formalism is the \emph{spin vertex} which is
the weak coupling analogy of the \emph{string vertex} in string field
theory.  We construct the spin vertex explicitly for all sectors at
the leading order using a set of bosonic and fermionic oscillators.
We prove that the vertex has trivial monodromy, or put in other words,
it is a Yangian invariant.  Since the monodromy of the vertex is the
product of the monodromies of the three states, the Yangian invariance
of the vertex implies an infinite exact symmetry for the three-point
function.  We conjecture that this infinite symmetry can be lifted to
any loop order.}

\begin{document}
\maketitle
\flushbottom

 \section{Introduction}
 \label{intro}

 The old idea of 't Hooft \cite{t1974planar} about the possibility of
 an exact correspondence between the multicolour QCD and some string
 theory has been realised two decades later for the simpler, conformal
 invariant "supersymmetric QCD", the maximally super-symmetric
 Yang-Mills theory
 \cite{Maldacena:1997re,Gubser:1998bc,Witten:1998qj}.  Even more
 excitingly, it has been discovered that the theory is likely to be
 integrable for all couplings.  After a crucial insight by Minahan and
 Zarembo \cite{MinahanZarembo}, and a great amount of collective work
 for one decade (see, for example the review \cite{Beisert-Rev}) it
 became clear that the spectral problem in AdS/CFT can be reformulated
 in terms of integrable spin chains, for which there exists a package
 of well developed mathematical methods originated from the Bethe
 Ansatz.  The spectral problem was formulated very elegantly in terms
 of the so called Quantum Spectral Curve \cite{2013arXiv1305.1939G}.
 We know in principle how to classify the eigenstates $|\CH_\a\> $ of
 the dilatation operator, which is represented by the Hamiltonian of
 the integrable spin chain, and to compute their correlation functions
 \be \< \CO_\a(x)\CO_\b(y)\> = {\d_{\a\b} \ \CN_a\over
 |x-y|^{\Delta_\a +\Delta_\b}}.  \ee
(It is convenient to use a normalisation in which the constants
$\CN_{\a} $ are given by the norms squared of the corresponding
on-shell Bethe states.)

 In the last few years the challenge moved from the spectral problem,
 i.e. the structure of the conformal dilatation operator by which is
 nowadays considered solved in principle, to the computation of the
 correlation functions, amplitudes and Wilson loops.  Understanding of
 the structures of these objects in the maximally supersymmetric
 theory would help devising efficient computation techniques for
 perturbative QCD. The structure of the interactions is encoded in the
 operator product expansion
 \be
 \la{OPE}
 \CO_\a(x)\CO_\b(y) \sim  C_{\a\b}^\g \CO_\g (y) \,
 |x-y|^{\Delta_\g - \Delta_\a-\Delta_\b},
 \ee
 or equivalently, in the three-point function of operators with given
 conformal weights:
  \be \< \CO_\a(x)\CO_\b(y)\CO_\g(z)\> = {C_{\a\b\g}\over
  |x-y|^{\Delta_\a +\Delta\b- \Delta_\g}
  |x-z|^{\Delta_\a+\Delta_\g-\Delta_\b}
  |y-z|^{\Delta_\b+\Delta_\g-\Delta_\a}}.  \ee
The two sets of constants are related by
 \be
 C_{\a\b}^\g = C_{\a\b\g}/ \CN_{\g }.
 \ee
 The structure constants involve trace operators with non-restricted
 lengths $L_1, L_2, L_3$.  There are two limits in which the problem
 can be approached by the available techniques, depending on the value
 of 't Hooft coupling $\l \sim g_{\text{YM}}^2 N_c$: that of extremely
 weak coupling $\l\to 0$, and that of extremely strong coupling,
 $\l\to\infty$.  Furthermore, the methods applied in each of the two
 limits depend on the values of the spin and the R-charges of the
 three operators.

At strong coupling, $\l\to\infty$, a general framework for computation
is given by string theory.  The methods depend on the type of
operators.  The {\it heavy } operators have large spin of R-charge and
correspond to classical strings moving in the AdS space.  In the case
of three heavy operators, the problem reduces to a generalisation of
the Plateau problem, namely to find a minimal surface embedded in the
AdS background and having prescribed singularities at three punctures.
The method to compute the classical action is based on the classical
integrability of the string sigma model \cite{Bena:2003wd}.  Each of
the three states represents a classical solution of the sigma model,
described by the spectral curve of the classical monodromy matrix.  A
major ingredient of the method is a condition on the monodromies
associated with the three punctures \cite{2011arXiv1109.6262J}.
Namely, the product of the three monodromies must be equal to one,
because the path can be contracted, but on other hand it gives a
non-trivial information about the solution, which is sufficient to
reconstruct the classical action.  In \cite{2011arXiv1109.6262J}, the
contribution from the AdS$_2$ part was evaluated for a string rotating
only on the sphere.  The full problem was solved in \cite{
Komatsu:su2}.  The case of three GKP strings, which requires also a
construction of the vertex operators, was solved in
\cite{Komatsu:3pt1, Komatsu:3pt2}, which led to a remarkably simple
formula in terms of contour integrals in the spectral plane.

 In the case of two heavy and one light operator, the methods are
 slightly different, but still based on the integrability.  This case
 was solved in \cite{Zarembo:2010ab,Costa:2010rz}.  The solution was
 recently given a major revision in \cite{Bajnok:2014sza}, where a
 missing modular integral was added.  This allowed to the authors of
 \cite{Bajnok:2014sza} to relate the computation with the form factor
 formalism \cite{Klose:2012ju,Bajnok:2014sza}, where the world-sheet
 integrability can be effectively used \cite{RoZoForm}.  Finally, the
 case of three short/medium operators has been worked out in
 \cite{Minahan:3ptshort2,Minahan:3ptshort,Minahan:2014usa}.

In the opposite limit $\l=0$ the gauge theory splits into a set of
non-interacting massless gaussian fields, a gauge boson, 6 scalars and
8 fermions, all in the adjoint representation of the gauge group
$U(N_c)$.  This limit is however not well defined because the spectrum
of the fields is highly degenerate: all traces of length $L$ have the
same dimension $\Delta = L$.  One can lift the degeneracy by switching
on temporarily the interaction, compute the one-loop eigenstates, and
then take again $\l=0$.  Then the operator $\CO_\a$ is described by an
on-shell Bethe state of the integrable spin chain\footnote{This is
well understood in the $su(2)$ and the $sl(2)$ sectors and not quite
well understood for the general eigenstates.} and is typically a sum
of terms the number of which grows factorially in the length of the
chain.  This is what makes the problem difficult.  Nevertheless, in
the $su(2)$ sector, a spectacular progress has been done in a series
of works \cite{EGSV,EGSV:Tailoring2,GSV,Pedro:sl23pt,Caetano:Fermionic}
where the
procedure called {\it Tailoring} was developed.  Tailoring reduces the
computation of the structure constant to the evaluation of the scalar
products of pairs of off-shell Bethe states representing segments of
spin chains.

There is no efficient way to compute such scalar products, except for
very short chains.  Fortunately, the structure constant in the
Escobedo-Gromov-Sever-Vieira (EGSV) configuration studied in
\cite{EGSV} can be expressed in terms of on-shell/off-shell scalar
products \cite{Omar, 2012arXiv1203.5621F}, for which there exists a
nice determinant formula \cite{slavnov-innerpr}.

The determinant representations allowed to generalise the results of
\cite{EGSV,EGSV:Tailoring2,GSV,Pedro:sl23pt} to the case of 3 non-BPS
fields in the EGSV configuration \cite{3pf-prl, SL}, to the case when
one of the fields is $su(3)$ type \cite{Foda:2013nua}, and extend the
result to the one-loop order
\cite{GV:quantumintegrability,Didina-Dunkl,GV,JKLS-fixing}.  One of
the exciting observantions made in \cite{JKLS-fixing} is the match (up
to some subtleties in choosing the integration contours) of the
one-loop structure constant with the $\l\to\infty$ result obtained in
\cite{Komatsu:su2}, in the Frolov-Tseytlin limit
\cite{Frolov-Tseytlin-l}.  Unfortunately the structure constant is
generically not a determinant and all these results cannot be used as
a basis for a systematic procedure.  In the same time, progress has
been made in the computation of the correlation functions in the
non-compact $sl(2)$ sector, based on very different techniques: the
method of separation of variables and the use of light ray
representation for the operators
\cite{Alday:higherspin3pt,2012arXiv1212.6563K,
2011JHEP09132G,2011arXiv1108.3557E,Eden3loops,Sobko:SoV}.

To summarise, in spite of these impressive achievements in various
particular cases, and in contrast with the spectral problem, there is
still no unified scheme for computing the correlation functions of
trace operators in $\CN=4$ SYM, which comprises all sectors at any
coupling.  The search of such a guiding principle based on the
integrability is the main subject of this paper.

There is no doubt that such a universal formalism should be based on
the notion of spin chain, which gives a description of the theory for
any coupling.  The spin chain can be also perceived as an integrable
discretisation of the string embedded in AdS$_5\times$S$^5$.  The
pertinence of such a picture comes from the fact that in the limit
$\l=0$ the string becomes tensionless and the indivisible units of the
string, the string bits, can be identified with the elementary
non-interacting fields in SYM.

We conjecture that the monodromy condition, which determines the
structure constant in the $\l \to \infty$, can be in principle
extended to any coupling down to $\l=0$.  Since we don't know the wave
functions for finite $\l$, the only check of this conjecture we can
afford at the moment is at $\l=0$.  In this limit the gauge theory
becomes a theory of 8 fermionic and 8 bosonic $N_c\times N_c$
non-interacting matrix fields.  In the string bit setup, we will
represent each of the fields by a pair of oscillators (one copy for
each site) and the colour indices will be taken care of by the
planarity constraints.  We are thus going to reformulate the
techniques used in different computations in gauge theory in a
language which is close to the formalism used in string theory, and
which we think will be adequate for the description of the
interactions.  A central concept of this formalism is the analogue of
the string field theory cubic vertex, which we call {\it spin vertex}.
This vertex should satisfy the monodromy condition or, put in other
words, should be Yangian invariant.  The Yangian invariance of the
spin vertex implies a condition on the correlation function of the
three operators.  If we restrict ourselves to the compact sector, this
can be formulated as a condition of the structure constant itself.

In this paper we consider only the tree level limit, but we hope that
the formalism can be extended for finite $\l$.  We first revisit the
analysis by Alday, David, Gava and Narain \cite{ADGN} of the
oscillator representation of the super-conformal algebra
$\mathfrak{psu}(2,2|4)$ based on its maximal compact subalgebra and
its relation with the standard representation, based on the stability
(little) group transforming the fields at $x=0$.  A key point in
\cite{ADGN} is that the non-unitary rotation $U$, which relates these
two representations, plays an important role in the computation of the
correlation functions and should be taken explicitly into account.  In
Section \ref{section:Oscill}, we improve on the ADGN construction of
the operator $U$, namely we show that this operator should act
nontrivially to the fermionic oscillators.  The full operator is a
product of a bosonic and a fermionic piece, $\U=U U_F$.

The Hilbert space for the states representing trace operators of
length $L$ is a tensor product of $L$ one-particle Fock spaces $V_+$
built on the Fock vacuum $\ket$.  The space $V_+$ includes a lowest
weight module of the superconformal algebra.  The correlation
functions contain an insertion of the operator $\U\U^\dag$, which
transforms the bosonic creation operators into annihilation operators.
Therefore we need a second, highest state module $V_-$, generated by
the action of the annihilation operators on the conjugate vacuum
$|\bar 0\> = \U^2 \ket$.  According to the formalism developed in
\cite{ADGN}, the two-point function of length-$L$ trace operators is a
bilinear form defined in the tensor product $V_+^{\otimes L}\otimes
V_+^{\otimes L}$, which can be translated, by the action of the
operator $\U$, into a bilinear form defined on $V_+^{\otimes L}\otimes
V_-^{\otimes L}$.  The main result of our work is is to redefine the
object introduced by ADGN \cite{ADGN} and which realizes the bilinear
map.  We are alternatively using two definitions,
\begin{align}
|V_{12}\rangle\in V_+^{\otimes L}\otimes V_+^{\otimes L}\;, \quad
|\CV_{12}\rangle= \U ^{2}_{(1)}|V_{12}\rangle \in V_-^{\otimes L}\otimes
V_+^{\otimes L}
\end{align}
The object $|\CV_{12}\rangle$, which we will refer to as 2-vertex, is
locally invariant under the super-conformal algebra,
\begin{align}
\label{losym}
\left( E_s^{AB(1)}+ E_s^{AB(2)}\right)
|\CV_{12}\rangle=0\;, \qquad s=1,\ldots, L
\end{align}
where $E^{AB(k)}_s$ are the generators of
$\mathfrak{u}(2,2|4)$ at site $s$ on chain $(k)$.  The
invariant $|\CV_{12}\rangle$ enters the expression of the two point
function as
\begin{align}
\label{eq1}
\langle {\cal O}_2(y) {\cal O}_1(x) \rangle = \langle \CV_{12}|\,
e^{i( L^+_1x+ L^+_2 y)} | \CO_2\rangle \otimes |{\cal O}_1\rangle \,.
\end{align}
with $ L^+_k$ generators in the oscillator representation associated
to the momentum operator.

\begin{figure}[h!]
\vskip 0.9cm
 \begin{center}
 \includegraphics[scale=0.55]{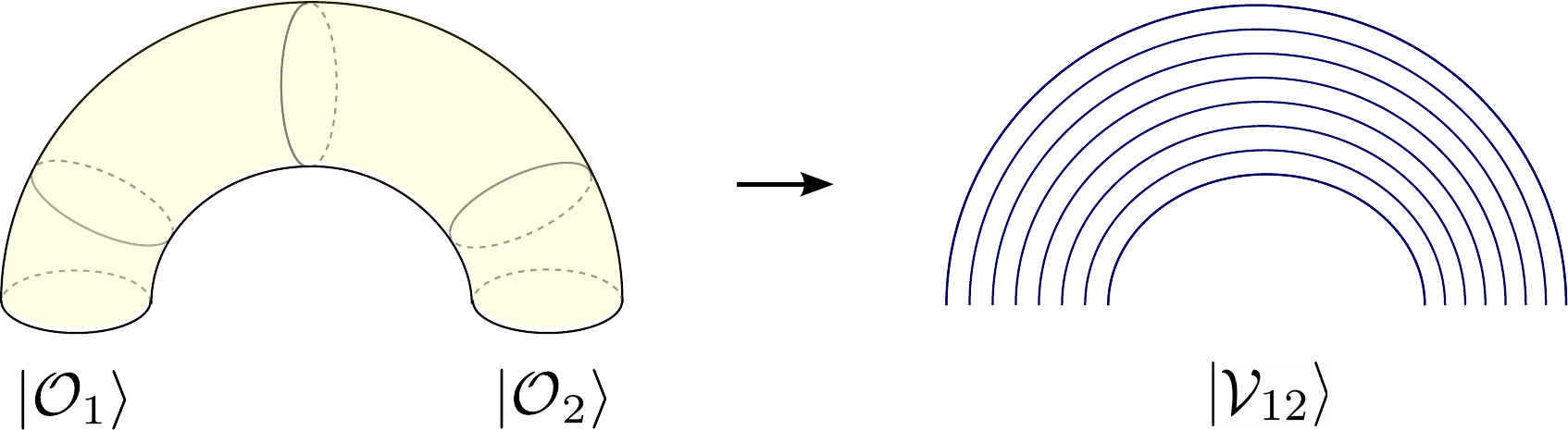}
\end{center}
 \vskip - 0.5cm
 \caption{ \label{figone} The two point correlation function and
 $|\CV_{12}\rangle$}
\end{figure}
\begin{figure}[h!]
  \begin{center}
  \vskip 0.7cm
  \includegraphics[scale=0.70]{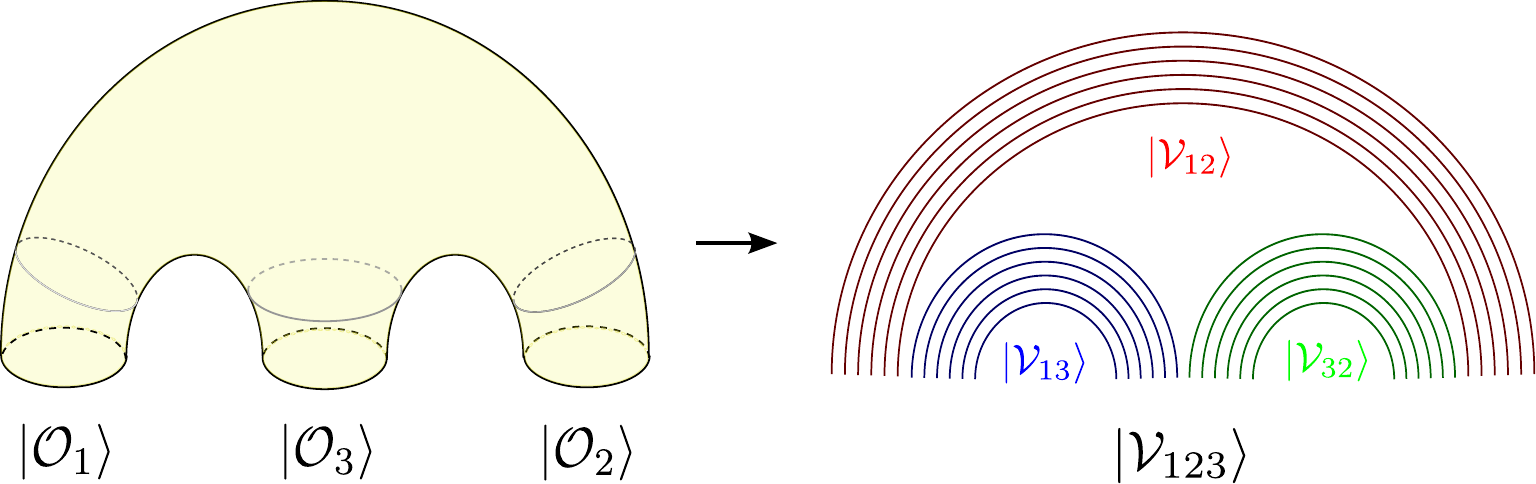}
  \end{center}
  \vskip - 0.9cm
   \caption{ \label{figtwo}  The three point correlation function and
   $|\CV_{123}\rangle$}
     \vskip 0.6cm
\end{figure}
The same strategy can be used to reformulate the three point function,
\begin{align}
\label{eq2}
\langle {\cal O}_2(y) {\cal O}_3(z) {\cal O}_1(x) \rangle = \langle
\CV_{123}|\, e^{i( L^+_1x+ L^+_2 y+ L^+_3z)} |\CO_2\rangle \otimes |
\CO_3\rangle \otimes |{\cal O}_1\rangle \,,
\end{align}
using the three-point invariant $ |\CV_{123}\rangle$ defined, at tree
level, as
\begin{align}
 |\CV_{123}\rangle=|\CV_{12}\rangle\otimes |\CV_{13}\rangle\otimes
 |\CV_{32}\rangle\;.
 \end{align}
The objects entering the correlation functions are schematically
depicted in figures \ref{figone} and \ref{figtwo}.  Such an
interpretation of the correlation functions is close in spirit to the
construction in \cite{Koch:2014nka}\footnote{We thank S. Komatsu for
bringing this paper to our attention.}, but it is also heavily
inspired from the ideas in string field theory, where the object
similar to $ |\CV_{123}\rangle$ is the string vertex
\cite{Spradlin:SFT2,Spradlin:SFT1,Pankiewicz:2002tg,
Shimada:2004sw,Dobashi:Resolving, Klose:lightcone3pt,Shulgin-Zayakin}.

\begin{figure}
         \centering
	 \begin{minipage}[t]{0.6\linewidth}
            \centering
            \includegraphics[width=6.2 cm]{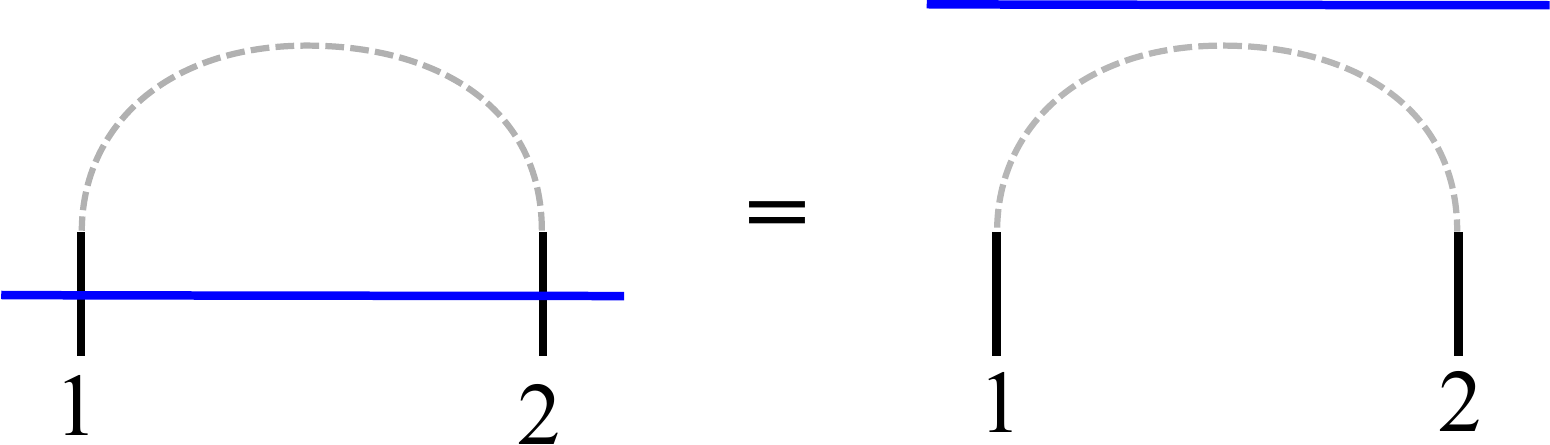}
\caption{ \small The basic relation of vertex operator, the two states
connected by the dashed line is identified.  }
  \label{basis}
         \end{minipage}
                    \end{figure}

\begin{figure}[h!]
\begin{center}
\bigskip \bigskip
\includegraphics[scale=0.32]{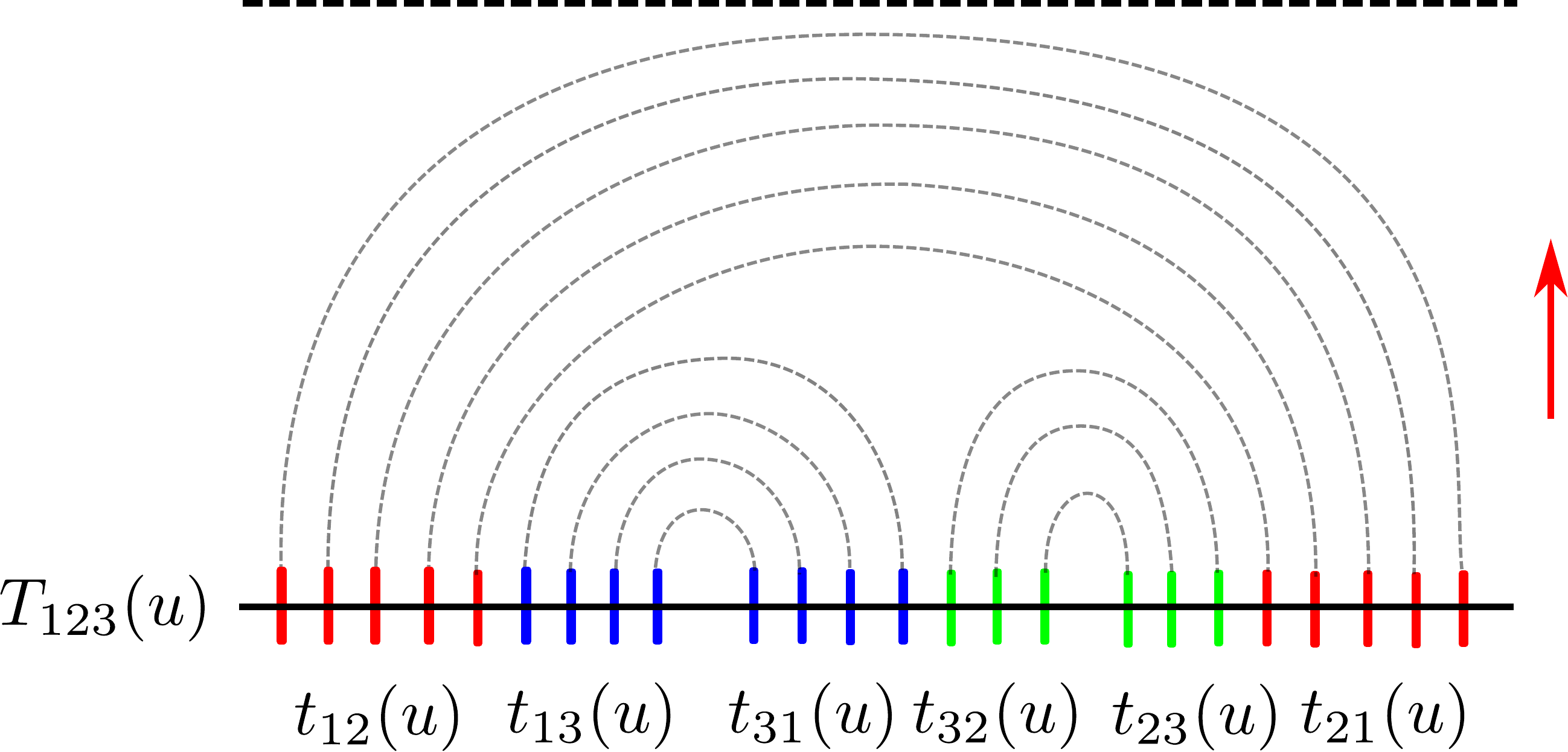}
\caption{ \label{3Vshift} The three-point spin vertex and monodromy
condition.}
\end{center}
\end{figure}

Compared to the previous works, the step forward we take here is to
reformulate the local symmetry (\ref{losym}) as a non-local symmetry
realized by the Yangian.  Of course, our aim is to reformulate the
symmetry conditions for the three point functions at arbitrary
coupling in terms of integrability.  Here, we show that at the tree
level the spin vertex $ |\CV_{123}\rangle$ is a Yangian invariant,
\begin{align}
T_{123}(u) |\CV_{123}\rangle= |\CV_{123}\rangle\;,
 \end{align}
with the monodromy matrix $T^{(123)}(u)$ built from the pieces of the
three chains, as shown in figure \ref{3Vshift},
\begin{align}
\label{tpvmono}
T_{123}(u)= t^{(12)}(u)\,  t^{(13)}(u)\,  t^{(31)}(u) \, t^{(32)}(u)\,
t^{(23)}(u)
\, t^{(21)}(u)\;.
\end{align}
The building block used for the monodromy condition is property of the
two-site vertex $|\CV_{12}\rangle$ carrying on the sites $1$ and $2$
the physical representation, as represented schematically in figure
\ref{basis},
  \begin{align}
\label{idvert}
R_{01}(u)R_{02} (u) |\CV_{12}\rangle =|\CV_{12}\rangle\;.
\end{align}
This relation can be traced back to the unitarity property of the $R$
matrix, $R_{01}(u)R_{01}(-u)=1$, plus a version of the crossing
relation mediated by the vertex.  Let us mention that the specific
form of the monodromy property (\ref{idvert}) concerns the full
$\mathfrak{psu}(2,2|4)$ R matrix and it changes when reduced to
particular subsectors.  The integrable structure displayed by the
vertex is instrumental in computing even the tree-level correlation
functions \cite{ShotaNew}, some of which were known previously.  We
think that the integrable structure will be maintained at higher
loops, and that the integrability constraints combined with few
general constraints will be sufficient to determine the three point
function, very much as the integrability constraints were sufficient
to determine the spectrum of anomalous dimensions \cite{Beisert-Rev}.

The structure of the paper is as follows: in section
\ref{section:Oscill} we are reviewing the oscillator representation
for the tree level $\mathfrak{psu}(2,2|4)$ algebra, as well as the
ADGN approach to computing the correlation functions using the Fock
space representation and the vertex.  In section \ref{section:Spinv}
we construct the spin vertex at tree level and we characterize its
properties, in particular how it flips outgoing states into incoming
states.  The subsection \ref{subsection:EGSV} shows how to reduce the
computation of correlation functions in the $\mathfrak{so}(6)$ sector
to overlaps, and how to retrieve the results obtains by EGSV
\cite{EGSV}.  In section \ref{section:Mono} we formulate the monodromy
condition and verify that it is satisfied for the auxiliary space in
the defining representation.  We end with conclusions and some
comments about the extension of the results at higher loops.

{\it Note:} We acknowledge that a part of the subjects discussed in
this paper is also investigated independently in the paper by Y.
Kazama, S. Komatsu and T. Nishimura \cite{Shota1:Yangian}.  Partial
results of the two groups were presented at the APCTP workshop in
Pohang \cite{ShotaPohang,DidinaPohang}.

\section{Oscillator representation and the free $\CN=4$ SYM }
\label{section:Oscill}

In determining the spectrum, the spin chain representation of the
dilation operator was very important.  This representation can be
easily understood using the oscillator representation of the algebra
$\mathfrak{psu}(2,2|4)$ \cite{Gunaydin1, Gunaydin2, Beisert-thesis}.
The oscillator representation, valid for the free field theory, is a
good starting point for setting up the perturbation theory.  The same
representation is also useful in computing the correlation functions,
since our aim is to reduce the computation of structure constants to
the evaluation of overlaps of wave functions of the spin chains.  In
this section we are reviewing the link between the oscillator
representation of $\mathfrak{psu}(2,2|4)$ and the standard unitary
presentation of the super-conformal group, link which is explained at
length in the reference \cite{ADGN}.  We refer to this article for
further details.

Let us first discuss the oscillator representation of the compact
version of $\mathfrak{psu}(2,2|4)$, $\mathfrak{psu}(4|4)$.  It uses
four copies of bosonic oscillators, $a_i, b_i,\ i=1,2$ and four copies
of fermionic oscillators, $c_k,\ k=1,\ldots,4$,
\begin{align}
[a_i,a_j^\dagger]=\delta_{ij}\;,\quad
[b_i,b_j^\dagger]=\delta_{ij}\;,\quad
\{c_k,c_l^\dagger\}=\delta_{kl}\;, \quad i,j=1,2\;, \quad
k,l=1,\ldots,4\;.
\end{align}
We organize the oscillators in a eight-dimensional vector
\begin{align}
\phi=(\begin{array}{ccc}a_i & b_i & c_k\end{array})
\end{align}
such that the generators of $\mathfrak{u}(4|4)$ can be written as
\begin{align}
 E^{AB}_{\rm compact}=\phi^{A\dagger}\phi^B \quad {\rm with} \quad
 E^{AB\dagger}_{\rm compact}= E^{BA}_{\rm compact}\;.
\end{align}
It is straightforward to check that they satisfy the commutation
relations of the $\mathfrak{u}(4|4)$ algebra,
\begin{align}
\label{commrelsu}
[ E^{AB},
E^{CD}]=\delta^{BC}E^{AD}-(-1)^{(|A|+|B|)(|C|+|D|)}\delta^{AD}E^{CB}\;,
\end{align}
with $[\cdot,\cdot]$ meaning commutator or anti-commutator, depending
on the grading of the generators, and the grading is $|A|=0,1$ for
bosonic and fermionic indices respectively.  The non-compact form
$\mathfrak{u}(2,2|4)$ can be obtained after a particle-hole
transformation for one group of bosonic oscillators, say $b$,
\begin{align}
 E^{AB}= E^{AB}_{\rm compact} (b\to -b^\dagger, b^\dagger\to b) \; .
\end{align}
The commutation relations (\ref{commrelsu}) are preserved by the
particle-hole transformation, but the Hermitian conjugate of the
generators are now
\begin{align}
\ E^{AB\dagger}= \gamma\, E^{BA}\gamma\;, \qquad \gamma={\rm
diag}(1_2,-1_2,1_4)\;.
\end{align}
Sometimes, for the sake of symmetry, it is convenient to perform also
a particle-hole transformation of the fermionic oscillators
\begin{align}
d_i=c_{i+2}^\dag\;, \qquad d_i^\dag=c_{i+2}\, \quad i=1,2\;.
\end{align}
Unlike the bosonic particle-hole transformation, the fermionic one is
unitary and therefore it does not change the real form of the algebra.
We will use alternatively the two notations.  The Lie-algebra
generators are expressed in terms of these oscillators as
\begin{align}
\la{EanFock} E^{AB}= \bar \psi^{A} \psi^B\;,
 \end{align}
with
\be \la{defpsi} \psi=(\begin{array}{cccc}a_i & \ - b^\dag _i &\ c_i &
\ d^\dag_i \end{array}), \quad \bar \psi =
\psi^\dag \g=(\begin{array}{cccc}a_i ^\dag & \ b _i &\ c_i ^\dag & \ d _i
\end{array}).  \ee
The projective condition in $\mathfrak{psu}(2,2|4)$ is obtained by
imposing that the identity generator $\sum_A E^{AA}=\bar \psi \psi $,
a central charge of the algebra, is zero,
\begin{align}
\label{ccrestriction}
\sum_A
E^{AA}=\sum_{i=1,2}(N_{a_i}-N_{b_i}+N_{c_i}+N_{c_{i+2}}-1)
=\sum_{i=1,2}(N_{a_i}-N_{b_i}+N_{c_i}-N_{d_i})=0\;,
\end{align}
\vskip-10pt \noindent where $N_a,\ N_b,\ N_c,\ N_d$ are the number of
the respective types of bosons and fermions in the two types of
representations.  The above condition selects two types of modules,
lowest weight $V_+$ and highest weight $V_-$, built upon two vacua
$|0\rangle$ and $|\bar 0\rangle$ respectively, dual to each other
\begin{align}
|0\rangle=|0\rangle_B\otimes |0\rangle_F\;, &\qquad |\bar
0\rangle=|\bar 0\rangle_B\otimes |\bar 0\rangle_F\;,\\
\nonumber (a_i,b_i,c_i,d_i)| 0\rangle=0\;, &\qquad
(a_i^\dag,b_i^\dag,c_i^\dag,d_i^\dag)|\bar 0\rangle=0\;, \quad
i=1,2\;.
\end{align}
It is worth mentioning that the particle-hole transformation
$(a_i,b_i,c_i,d_i)\to (-a_i^\dag,-b_i^\dag,c_i^\dag,d_i^\dag)$ and
$(a_i^\dag,b_i^\dag,c_i^\dag,d_i^\dag)\to (a_i,b_i,c_i,d_i)$ helps
defining another copy of the $\mathfrak{psu}(2,2|4)$ generators that
act naturally in the dual module $V_-$ and which are the particle-hole
transformed of the generators (\ref{EanFock}).  The new generators can
be shown to be equal to
\begin{align}
\label{conjgen}
\bar
E^{AB}=-(-1)^{|B|}\psi^A\bar\psi^B&
=-(-1)^{|B|+|A||B|}E^{BA}-(-1)^{|B|}\delta^{BA}\nonumber
\\
&=-(E^{AB}+ (-1)^{|B|}\delta^{BA})^t\;,
\end{align}
where the index $^t$ stands for the super transposition.

Let us now concentrate on the conformal subalgebra in four dimensions
$\mathfrak{so}(2,4)\simeq \mathfrak{su}(2,2)$.  In the above
oscillator representation, there is a natural grading with respect to
the maximal compact subalgebra $\mathfrak{u}(1) \otimes
\mathfrak{su}(2)\otimes \mathfrak{su}(2)$.  The grading is given by
the value of the $\mathfrak{u}(1)$ generator $E$
\begin{align}
 [ E, L^\pm]&=\pm L^\pm\;, \qquad [ E,L^0]=0\;.\nonumber
\end{align}
In other words, the generators $L^0$ from the maximal compact subgroup
preserve the number of bosons, while $L^\pm$ increase or decrease the
number of bosons by 2.  We are going to use later the explicit
representation of these operators in terms of oscillators,
\begin{align}
 E=1&+\hf (N_a+N_b)=\hf (a^\dagger a +b b^\dagger)\;,\\
L_\mu^+&=-a^\dagger\bar \sigma_\mu b^\dagger\;, \quad L_\mu^-=b
\sigma_\mu a\;,
\end{align}
with $\sigma_\mu=(-1,\vec \sigma)$, and $\bar \sigma_\mu=(-1,-\vec
\sigma)$ and summation over indices of the bosonic operators is
understood.  For the $R$ charge sector, the generators are those of
the $\mathfrak{su}(4)$ algebra
\begin{align}
{\rm R}^{kl}=c_k^\dagger c_l-\frac{1}{4}\delta_{kl}\,c^\dagger c\;.
\end{align}

We will now identify the above generators with the standard
presentation of the conformal group, which is the group of rotations
in a six-dimension space with signature $\eta_{PQ}={\rm
diag}(-++++-)$.  We adopt the same convention as in \cite{ADGN} and
call the directions in the six-dimensional space $P,Q=0,1,2,3,5,6$,
with the first four directions corresponding to the Minkovski space,
$\mu, \nu=0,1,2,3$ .  The commutation relation are
\begin{align}
 [M_{PQ},M_{RS}]=i(\eta_{QR}M_{PS}-\eta_{PR}M_{QS}-\eta_{QS}M_{PR}
 +\eta_{PS}M_{QR})\;,
\end{align}
and the identification of the generators for translations $P_\mu$,
special conformal transformations $K_\mu$ and dilatation $D$ is made
as
\begin{align}
 P_\mu=M_{\mu 6}+M_{\mu5}\;, \qquad K_\mu=M_{\mu 6}-M_{\mu5}\;, \qquad
 D=-M_{56}\;.
 \end{align}
 On the other hand, the $\mathfrak{u}(1)$ generator in the oscillator
 representation $E$ is given by
\begin{align}
 E=M_{06}=\hf (P_0+K_0)\;.
 \end{align}
 The authors of \cite{ADGN} suggested that the oscillator
 representation and the standard representation above can be related
 by a transformation which exchanges the two directions with opposite
 signature $0$ and $5$, that is a rotation with an imaginary angle
 $-i\pi/2$ in the plane $05$,
\begin{align}
 U=\exp -\frac{\pi}{2} M_{05}=\exp- \frac{\pi}{4} (P_0-K_0)\;,
 \end{align}
 and its action translates into
\begin{align}
 U^{-1}K_\mu U=L_\mu^-\;, \quad U^{-1}P_\mu U=L_\mu^+\;, \quad U^{-1}D
 U=iE\;,
 \end{align}
\begin{align}
 L_0^+-L_0^-=U^{-1}(P_0-K_0) U=P_0-K_0\;,
 \end{align}
 which helps make contact between rotated and unrotated
 representations.  The transformation implemented by $U$ is similar to
 the so-called mirror transformation in two-dimensional field
 theories, including the AdS/CFT sigma model.  From the space-time
 interpretation, it is obvious that this transformation should obey
 $U^4=1$ except on spinors, and that $U^2$ is a kind of PT
 transformation which changes the sign of both 0 and 5 coordinates,
\begin{align}
 \label{crossingrel}
 U^{-2}D U^2=-D\;, \qquad U^{-2}E U^2=-E \;.
 \end{align}
 This relation is purely algebraic and it holds at any loop level, as
 it can be seen putting
\begin{align}
  \nonumber U_t=\exp \frac{t}{2} (P_0-K_0)\;, \qquad
  D_t=U_t^{-1}DU_t\;, \qquad E_t=U_t^{-1}EU_t\;.
   \end{align}
Taking the derivative with respect to $t$ and using
$E=\frac{1}{2}(P_0+K_0)$ and $D=\frac{i}{2}[P_0,K_0]$ and the
commutation relations of the conformal algebra, we get $\partial_t
E_t=iD_t$ and $\partial_t D_t=iE_t$, which is solved by
\begin{align}
 D_t=D\cos t +iE\sin t\;, \qquad E_t=E\cos t +iD\sin t\;.
 \end{align}
At tree level, the oscillator representation of the hermitian operator
$U$ is
\begin{align}
    U= \exp -\frac{\pi}{4}\sum_{i=1,2}(a^\dagger_ib^\dagger_i+
    a_ib_i)\;, \qquad U^\dag= U \;.
 \end{align}
By inspection, using the oscillator representation, we find that
\begin{align}
 \label{crossingrell}
   U^{2} L_\mu^+ U^{-2}=-&b \bar \sigma_\mu a\;, \qquad U^{2} L_\mu^-
   U^{-2}=a^\dagger \sigma_\mu b^\dagger\;, \quad {\rm or}\\
   U^{2} L_0^\pm U^{-2}&=- L_0^\mp \;, \qquad U^{2} L_m^\pm U^{-2}=
   L_m^\mp \nonumber\;,\\
    U^{2} P_0 U^{-2}&=- K_0 \;, \qquad U^{2} P_m U^{-2}= K_m\;.
    \nonumber
 \end{align}
These relations can be derived using the action of the operator $U$ on
the oscillators, in particular
\begin{align}
\label{ucarreosc}
   U^{2}a U^{-2}=b^\dagger\;, \quad U^{2}a^\dag U^{-2}=-b\;,\quad
   U^{2}b\, U^{-2}=a^\dagger\;, \quad U^{2}b^\dag U^{-2}=-a\;.
 \end{align}
From here we conclude also that the transformation $U^2$ sends the
bosonic Fock vacuum $ | 0\rangle_B$ into the dual vacuum $ |\bar
0\rangle_B$,
\begin{align}
 |\bar 0\rangle_B = U^2| 0\rangle_B\;, \qquad a^\dagger , b^\dagger
 |\bar 0\rangle_B =0\;,
  \end{align}
 therefore mapping the lowest weight module $V_+$ to the highest
 weight one $V_-$ and back,
\begin{align}
 V_+ \ {\buildrel U^2 \over \longleftrightarrow} \ V_-\;.
  \end{align}
  Given the relation (\ref{crossingrel}), we may conclude that the
  positive energy states belong to $V_+$ and the negative energy ones
  belong to $V_-$, where the term of energy refers to the eigenvalues
  of the operater $E$.  We note from the relations (\ref{ucarreosc})
  that
\begin{align}
\label{uquatreosc}
  U^{-4} x\, U^4= -x , \qquad x=a_{i},a_i^\dagger,
  b_{i},b_i^\dagger\;,
 \end{align}
 which does not pose a problem for the generators which are quadratic
 in the bosons or in the fermions, but it changes the sign of the odd
 generators of the super-conformal group, which transform in the
 spinorial representations of both $\mathfrak{so}(6)$ and
 $\mathfrak{so}(4,2)$.  Therefore, we may supplement the operator $ U$
 with a fermionic counterpart $ \C$, such that $\U=( U \C)^4$ will
 change the sign of the fermions as well,
\begin{align}
\label{uwithf}
   \C= \exp -\frac{\pi}{4}\sum_{i=1,2}(c_i^\dagger d_{i}^\dagger+ c_i
   d_{i})\;,\qquad \C^\dag= \C^{-1}\;.
 \end{align}
 In other words, the non-unitary rotation in space-time is
 supplemented by an unitary rotation in the $R$ charge sector, which
 is the product of two $\mathfrak{su}(2)$ rotations that will be
 called later $\mathfrak{su}(2)_L$ and $\mathfrak{su}(2)_R$.
 The action of the transformation on the fermionic oscillators is
\begin{align}
\label{ucarreoscf}
    \C^{2} c_i\, \C^{-2} = d_{i}^\dag, \quad & \C^{2} c_i^\dag\,
    \C^{-2}= d_{i}, \quad \C^{2} d_{i}\, \C^{-2} = - c_i^\dag, \quad
    \C^{2}d^\dag_{i} \, \C^{-2} = - c_i\nonumber \\
    & \C^{4} x\, \C^{-4}= -x , \qquad x=c_{i}, c_i^\dagger\, d_i,
    d_i^\dag\,\\
   &\qquad \C^{-2}\equiv \sigma=-\sigma_{2,L}\, \sigma_{2,R} \nonumber
    \end{align}
   and it also transforms the fermionic vacuum into its conjugate,
\begin{align}
 |\bar 0\rangle_F = \C^2| 0\rangle_F \equiv \C^2 |Z\rangle=c_1^\dag
 d_1^\dag c_2^\dag d_2^\dag |0\rangle_F \equiv |\bar Z\rangle\;.
 \end{align}
 Let us note that $ \C^2$, being a rotation, maps $V_\pm$ to
 themselves,
\begin{align}
 V_\pm \ {\buildrel \C^2 \over \longleftrightarrow} \ V_\pm\;.
  \end{align}

\subsection{Oscillator representation and the correlation functions}

We have now the necessary ingredients to present the dictionary
between the gauge invariant operators in the conformal field theory
and the Fock space representation.  The gauge invariant operators we
will consider in the planar limit are the single traces on the gauge
group, or ``words'' made up from the ``letters'' which are the
fundamental fields of the theory -- and which were interpreted as
string bits in view of the gauge-string correspondence,
\begin{align}
 \CO(x)\sim \Tr (XXZY\Psi_i\ldots)(x)\;.
\end{align}
   When the gauge coupling constant is zero, these string bits are
   independent and each of them is in a state corresponding to the
   $\mathfrak{psu}(2,2|4)$ representation described above.  Gauge
   invariant operators can then be represented by elements in the
   tensor product of the individual string bits.  In the spin chain
   representation, string bits are the sites of the spin chain, and we
   will have to introduce a copy of oscillators on each site $s$,
\begin{align}
\psi_s=\left(\begin{array}{cccc}a_{i,s} & -b^\dagger_{i,s} & c_{i,s}&
d^\dag_{i, s}\end{array}\right)\;, \quad s=1,\ldots,L\;
\end{align}
acting in the tensor product of individual sites $V_\pm^{\otimes
L}=V_{1,\pm}\otimes \dots \otimes V_{L,\pm}$.  In the non-interacting
gauge theory, the oscillator representation of the super-conformal
group generators will be
\begin{align}
  E^{AB}= \sum_{s=1}^L E^{AB}_s \;, \quad \U = \U_1\otimes\dots
  \otimes \U_L,
\end{align}
while the radiative correction will introduce interaction between the
string bits, or sites.  The space of conformal primary operators
${\cal O}(x)$ situated at $x=0$ is selected by the condition
\begin{align}
K_\mu{\cal O}(0)=0\;.
\end{align}
On the other hand, we have for the Fock vacuum $|0\rangle =
\ket_1\otimes \dots \otimes\ket_L$
\begin{align}
L^-_\mu|0\rangle=0\;, \quad {\rm hence} \quad K_\mu U |0\rangle=0\;.
\end{align}
Similarly, following \cite{ADGN}, we can relate the space of conformal
primary operators with the space of Fock states $|{\cal O}\rangle$
annihilated by the $L^-_\mu$ operator,
\begin{align}
  L^-_\mu|{\cal O}\rangle=0\;, \quad \Rightarrow \quad K_\mu U |{\cal
  O}\rangle=0\;.
\end{align}
Translating the operators to a different space-time point can be done
with the help of the momentum operator,
\begin{align}
{\cal O}(x)=e^{iP x}{\cal O}(0)e^{-iP x}\;,
\end{align}
with corresponding Fock space representative
\begin{align}
e^{i P x} U |{\cal O}\rangle\;.
\end{align}
For the operators $\CO$ with definite conformal dimension $\Delta$ we
have\footnote{This equation might seem paradoxical, since the
dilatation operator is hermitian and it should have real eigenvalues.
However, the state $U |{\cal O}\rangle$ has infinite norm and
therefore $i\Delta$ does not belong to the spectrum.}
\begin{align}
  D\, U |{\cal O}\rangle=i\, U E |{\cal O}\rangle=i\,\Delta \, U
  |{\cal O}\rangle \;,
\end{align}
so that
\begin{align}
e^{i D \ln \Lambda} U |{\cal O}\rangle=\Lambda^{-\Delta}\, U |{\cal
O}\rangle \;.
\end{align}
A similar identification holds between the bra states and the
hermitian conjugates of the operators,
\begin{align}
{\cal O}^\dagger(x)\quad \longleftrightarrow \quad \langle {\cal
O}|U^\dag e^{-iP x}= \langle {\cal O}|U^\dag e^{-iP x} \;.
\end{align}
This mapping was used by the authors of \cite{ADGN} to write the two
point function in terms of the Fock space representation,
\begin{align}
  \label{maptpf}
\langle {\cal O}_2^\dagger(y) {\cal O}_1(x) \rangle = \langle {\cal
O}_2| U^\dag e^{i P(x- y)} U |{\cal O}_1\rangle= \langle {\cal O}_2|
U^2 e^{i L^+(x- y)} |{\cal O}_1\rangle \; .
\end{align}
The authors of \cite{ADGN} also verified that if ${\cal O}$ is any
elementary field, for example $Z$, the tree-level
representation of the operators in the Fock space gives the desired
result of the Wick contraction
 %
\begin{align}
 \label{corrtree}
\langle \bar Z(x) {Z}(y) \rangle = \langle Z|U^2 e^{iL^+(y- x)} |Z\rangle
=\frac{ \langle Z|Z\rangle}{(x-y)^2}=\frac{1}{(x-y)^2}\;.
\end{align}
To get the next to the last equality sign, one has to use $L^+_\mu
=-a^\dagger \bar \sigma_\mu b^\dagger$ and, as suggested in \cite{ADGN},
to regularise $U^2$
as $U ^2= \lim_{t\to -\pi/2}U_{t}$, with $U_t$ given by
\begin{align}
\label{norf}
U_t=\exp t(a^\dag b^\dag+ba)=\exp (a^\dag b^\dag \tan t)\exp (-(a^\dag
a+bb^\dag)\ln \cos t)\exp (ab \tan t)\;.
\end{align}
(We give the details in Appendix \ref{AppendixA}.)  In fact, the
relation above should hold at higher loop as well,
\begin{align}
U_t= \exp -t(L_0^+-L_0^-)=\exp (- L_0^+ \tan t)\exp (-2E\ln \cos
t)\exp (L_0^- \tan t)\;,
\end{align}
since the commutation relations $ [ E, L_0^\pm]=\pm L_0^\pm$ and
$2E=[L_0^+,L_0^-]$ are the same at any coupling.  The last equality
sign in (\ref{corrtree}) amounts to computing the overlap for the
vacuum state,
\begin{align}
\langle Z|Z\rangle=1\;, \qquad |Z\rangle = |0\rangle\;.
\end{align}
A similar representation can be used for the special case of the
extremal three point function,\footnote{This example is only
illustrative since we are not computing an extremal correlation
function even at tree level, because of the mixing of single-trace and
double-trace states \cite{Rastelli-CorE}.  } when the length of the
first chain equal the sum of the lengths of the second and the third,
$L_1=L_2+L_3$,
%
\begin{align}
\langle {\cal O}_2^\dagger(y) {\cal O}_3^\dagger(z) {\cal O}_1(x)
\rangle_{\rm ext} = \langle {\cal O}_2|\otimes \langle {\cal O}_3|
U_{2} U_{3} \, e^{i P_{1}x} e^{-i P_{3}z} e^{-i P_{2}y} U_{2} |{\cal
O}_1\rangle_{\rm ext} \;,
\end{align}
where the index on the operators shows now the space on which they
act.  At tree level for the extremal correlator $U_{1}=U_{2}U_{3}$ and
$P_{1}=P_{2}+P_{3}$.  We conclude from the above that the correlators in the
Fock space representation involve a pairing between states in the
$V_+$ module in the ket states and the $V_-$ module in the bra states.

\subsection{The necessity of the spin vertex}

The Fock space representation is easily understood for the two point
function and the extremal three point function, where at weak coupling
the number of sites (string bits) is conserved from the bra to the ket
states.  The situation is more subtle for non-extremal correlation
functions, where the chains are splitting and joining, and some pieces
of the chains have to be flipped (see e.g. \cite{EGSV}) in order to contract them with pieces
of a different chain.  Let us now interpret the two point correlator
in (\ref{maptpf}) in a slightly different manner, considering now that
both operators act on the left Fock space.  To do this, we need a
mapping from a left state $\langle{\cal O}|$, to a right state $| \bO
\rangle$, which will be done via a specially prepared state $\langle
V_{12}|$ which lives in the tensor product of two chains,
 %
\begin{align}
  \label{mapdualstate}
  \; ^{(1)}\langle{\cal O}|= \langle V_{12}|\,\sigma^{(1)}\, |\bO
  \rangle^{(2)}\;,
  \qquad \sigma\equiv U_{F}^{-2},
\end{align}
where we have added an index to the Fock spaces to emphasize that
$\langle{\cal O}|$ and $| \bO \rangle$ live in different modules
$(V_{+}^{\otimes L})^{(1)}$ and $(V_{+}^{\otimes L})^{(2)}$ intertwined by
$\langle V_{12}| $.
We will show in section \ref{subsec:barO} that
the state $| \bO\rangle$ is the flipped state with respect to
$\langle{\cal O}|$ in the sense of \cite{EGSV}, being different from $| {\cal O}\rangle$.  In this
language, the two point function is
\begin{align}
  \label{maptpfver}
\langle  {\cal O}_2^\dagger(y){\cal O}_1(x) \rangle
&= \langle V_{12}
| \,\U_{(1)}^{\dagger 2} \; e^{iL^+_{(1)}(x- y)} \, |{\bO}_2\rangle^{(2)}\otimes
|\CO_1\rangle^{(1)}
\no  \\
&  =\nonumber \langle
\CV_{12} | \, e^{iL^+_{(1)}(x- y)} \, |{\bO}_2\rangle^{(2)}\otimes
|\CO_1\rangle^{(1)}
\\
& = \langle \CV_{12} | \, e^{i[ L^+_{(1)}x +
L^+_{(2)}y ]} \, |{\bO}_2\rangle^{(2)}\otimes
|\CO_1\rangle^{(1)} \;,
\end{align}
where $\U^2=U^2\C^2$ and $\U^{\dag2}=U^2\C^{-2}$.  In the second  line we
have introduced the state
\begin{align}
 \label{symL}
\langle \CV_{12} | \equiv \langle V_{12} | \, \U_{(1)}^{\dag
2}\;,\qquad | \CV_{12} \rangle \in V_-^{\otimes L}\otimes V_+^{\otimes
L}\;,
 \end{align}
 and used the property which we will prove later
\begin{align}
 \label{symLbis}
 \langle V_{12} | \, \U_{(1)}^{\dag 2}\, (L_{(1)}^++L_{(2)}^+)\equiv \langle
 \CV_{12} | (L_{(1)}^++L_{(2)}^+)=0\;.
 \end{align}
The state $\langle \CV_{12}| $, or its conjugate $|\CV_{12}\rangle$,
should play the role of the vacuum state, in the sense that is has to
carry the same quantum numbers as the vacuum. It is clear that $|\CV_{12}\rangle$
cannot be the tensor product of the Fock space vacua of the two
chains.  At tree level, $|{\cal V}_{12}\rangle$ should provide the right Wick
contractions between the elementary fields in $\CO_2^\dagger$ and
$\CO_1$.  A similar relation holds for the extremal three
point function,
\begin{align}
  \label{mapttpfverext}
\langle{\cal O}^\dagger_2(y) {\cal O}^\dagger_3(z) {\cal O}_1(x)
\rangle _{\rm ext}& = \,_{\rm ext}\langle V_{123}|\, \U_{(1)}^{\dagger
2}\; e^{i[ L^+_{(1)}x+ L^+_{(2)} y+ L^+_{(3)}z]} |\bO_2\rangle
\otimes | \bO_3\rangle \otimes |{\cal O}_1\rangle \\
&= \,_{\rm ext}\langle \CV_{123}|\, e^{i[ L^+_{(1)}x+ L^+_{(2)} y+
L^+_{(3)}z]} |\bO_2\rangle \otimes | \bO_3\rangle \otimes |{\cal
O}_1\rangle \,, \nonumber
\end{align}
where the extremal vertex $ |\CV_{123}\rangle_{\rm ext} $ is built
from two pieces connecting each the operators $\CO_2$ and $\CO_3$ with
$\CO_1$,
\begin{align}
 |\CV_{123}\rangle_{\rm ext}=|\CV_{12}\rangle\otimes
 |\CV_{13}\rangle\;.
\end{align}
In this case, at tree level there are Wick contractions only between
the operators 1 and 2 and 1 and 3 and there are no contractions
between the operators 2 and 3.  At this point we are starting to see
that in the vertex formulation the operators can be treated more
democratically,
\begin{align}
  \label{demtpfext}
\langle {\cal O}_2(y)  {\cal O}_3(z) {\cal O}_1(x) \rangle _{\rm ext} =
\,_{\rm ext}\langle \CV_{123}|\, e^{i[ L^+_{(1)}x+ L^+_{(2)} y+
L^+_{(3)}z]} |\CO_2\rangle \otimes | \CO_3\rangle \otimes |{\cal
O}_1\rangle \;.
\end{align}
This helps to define the slightly more complicated case of a
non-extremal three point function, where the operators $\CO_2$ and
$\CO_3$ are also connected by Wick contractions.  At tree level, we
can split any of the operators $\CO_i$ into pieces $\CO_{ij}$ which are contracted to
pieces $\CO_{ji}$ of operator $\CO_j$.  At the level of the states we
have \footnote{The writing below is does not imply that the state
associated to the operator 3 is a product, just that it belongs to the
tensor product of the Fock spaces denoted by 31 and 32.}
\begin{align}
| \CO_1\rangle=| \CO_{13}\rangle\otimes| \CO_{12}\rangle\;,\\\nonumber
| \CO_2\rangle=| \CO_{21}\rangle\otimes| \CO_{23}\rangle\;,
\\\nonumber | \CO_3\rangle=| \CO_{32}\rangle\otimes|
\CO_{31}\rangle\;.
\end{align}
The non-extremal three point function, at tree level, can be then
written in the same way as non-extremal, but with another definition of the vertex
\begin{align}
  \label{mapttpfverbisinc}
\langle{\cal O}_2(y) {\cal O}_3(z)  {\cal O}_1(x) \rangle = \langle
\CV_{123}|\, e^{i[ L^+_{(1)}x+ L^+_{(2)} y+ L^+_{(3)}z]}
|\CO_2\rangle \otimes | \CO_3\rangle \otimes |{\cal O}_1\rangle \,,
\end{align}
with the vertex $|\CV_{123}\rangle$ built out as
\begin{align}
 |\CV_{123}\rangle&=|\CV_{12}\rangle\otimes |\CV_{13}\rangle\otimes
 |\CV_{32}\rangle=\U^2_{(12)}|V_{12}\rangle\otimes\U^2_{(13)}
 |V_{13}\rangle\otimes \U^2_{(32)} |V_{32}\rangle\;\;, \nonumber \\
    |\CV_{ij}\rangle &\in V_-^{\otimes L_{ij}}\otimes V_+^{\otimes
  L_{ji}}\;.
\end{align}
The construction of the states $|{\cal V}_{12}\rangle$ and $|{\cal V}_{123}\rangle$,
that we call the spin vertex (by abuse of language we will call
$|{\cal V}_{12}\rangle$ the two-vertex) is the main purpose of this work.

\section{The spin vertex at tree level}
\label{section:Spinv}

In this section we are defining the basic building blocks we need to
build the vertex at tree level.  The main object is the two-vertex
$|\CV_{12}\rangle$, which is an invariant of the
$\mathfrak{su}(2,2|4)$ algebra and which can be therefore used as a
``vacuum state'' in the tensor product of multiple Fock spaces when we
compute the correlation functions.

\subsection{Definition of the two-vertex }

We will concentrate first on the case of the two-vertex
$|\CV_{12}\rangle$ and infer the properties required such that
(\ref{maptpf}) and (\ref{maptpfver}) are identical.
 A construction of the vertex using the oscillator representation was
 given in \cite{ADGN}.  Here we give a slightly modified version of
 that construction\footnote{The main difference between our definition
 of the vertex and the one in \cite{ADGN} is that our vertex is
 neutral for the R-charges while theirs is not.}
\begin{align}
\label{Aver}
 |{\cal
 V}_{12}\rangle&\equiv\U_{(1)}^2|{V}_{12}\rangle
 \no
\\
 &=\U_{(1)}^2\exp\sum_{s=1}^L\sum_{i=1,2}
 \left(b^{(1)\dagger}_{i,s}a^{(2)\dagger}_{i,s}
 -a^{(1)\dagger}_{i,s}b^{(2)\dagger}_{i,s}
 -d^{(1)\dagger}_{i,s}c^{(2)\dagger}_{i,s}
 -c^{(1)\dagger}_{i,s}d^{(2)\dagger}_{i,s}\right)
 \nonumber |0\rangle^{(2)}\otimes |0\rangle^{(1)}\\
 &=\exp-\sum_{s=1}^L\sum_{i=1,2}\left(a^{(1)}_{i,s}
 a^{(2)\dagger}_{i,s}-b^{(1)}_{i,s}b^{(2)\dagger}
 _{i,s}+d^{(1)}_{i,s}d^{(2)\dagger}_{i,s}-c^{(1)}_{i,s}c^{(2)\dagger}_{i,s}
 \right) | 0\rangle^{(2)} \otimes | \bar 0\rangle^{(1)} \; ,
  \end{align}
 where the upper index on the oscillators indicates the Fock space
 where they act, and $|\bar 0\rangle=\U^2| 0\rangle$.  In order to
 mimic the planar contractions we revert the order of the tensor
 product in the second chain,
\begin{align}
   \label{revord}
 |0\rangle^{(2)}\otimes |0\rangle^{(1)}=
 \left( |0\rangle_L^{(2)}\otimes\dots\otimes
 |0\rangle_1^{(2)}\right)
\otimes  \left(
 |0\rangle_1^{(1)}\otimes \ldots \otimes|0\rangle_L^{(1)}
 \right) \;.
\end{align}
%
%
The vertex
(\ref{Aver}) can be expanded as
\begin{align}
  \label{vertesstates}
|\CV_{12}\rangle&=\sum_{N_a,N_b,N_c,N_d}|N_a,N_b,N_c,N_d\rangle^{(2)}\otimes
 |\bar N_a,\bar N_b,\bar
N_c,\bar N_d\rangle^{(1)}\\ \nonumber &=\sum_{N_a,N_b,N_c,N_d} |\bar N_a,\bar N_b,\bar
N_c,\bar N_d\rangle^{(2)}\otimes |N_a,N_b,N_c,N_d\rangle^{(1)}=(-1)^{F}|\CV_{21}\rangle
\
\;,
\end{align}
where $F=N_c+N_d$ is the number of fermions and
\begin{align}
| N_a,N_b, N_c,
N_d\rangle=&\frac{1}{\sqrt{N_a!\,N_b!}}
\prod_{k=1,2}
(d_k^\dag)^{N_{d_k}}(c_k^\dag)^{N_{c_k}}(b_k^\dag)^{N_{b_k}}(a_k^\dag)^{N_{a_k}}|0\rangle\;,
\\ \nonumber
|\bar N_a,\bar N_b,\bar N_c,\bar
N_d\rangle&=\frac{(-1)^{N_a+N_c}}{\sqrt{N_a!\,N_b!}}
\prod_{k=1,2}a_k^{N_{a_k}}b_k^{N_{b_k}}c_k^{N_{c_k}}d_k^{N_{d_k}}|\bar
0\rangle\;,
 \end{align}
with $N_{a}!\equiv N_{a_1}!\,N_{a_2}!$ and $N_{b}!\equiv N_{b_1}!\,N_{b_2}!$.  For the states containing fermions one should take care of signs, so the order on which the fermionic oscillators act is important. In the formulas above we take the convention that the oscillators act in opposite order on the two chains.  One
can easily project the vertex in (\ref{vertesstates}) on the states
obeying $N_a-N_b+N_c-N_d=0$.
The second line in  (\ref{vertesstates}) can be proven using
\begin{align}
 \U_{(1)}|V_{12}\rangle=\U_{(2)}^{-1}|V_{12}\rangle\;,\qquad  \U_{(1)}|\CV_{12}\rangle=\U_{(2)}^{-1}|\CV_{12}\rangle\;,
 \end{align}
which will can be shown using the properties (\ref{exchangeosc}) below.
From the oscillator expansion (\ref{vertesstates}) it can be readily seen that
\begin{align}
  \label{vertexid}
\langle \CV_{31}|\CV_{12}\rangle=\sum_{N_a,N_b,N_c,N_d}|N_a,N_b,N_c,N_d\rangle^{(2)}\,^{(3)} \langle N_a,N_b,N_c,N_d | ={\bf 1}_{23}\;,
\end{align}
with ${\bf 1}_{23}$ identifying the spaces $3$ and $2$.

In order for the vertex $|\CV_{12}\rangle$ to reproduce the right two point functions of the operators in ${\cal N}=4$ SYM, it  has to contain, for each site $s$, the
``lowest weight'' state $ |Z\rangle_s\otimes|\bar Z\rangle_s$, as well
as the other combinations, $ |a\rangle_s\otimes|\bar a\rangle_s$ with
$a=Z,X,Y,\bar Z,\bar X, \bar Y$, plus the fermions, etc.
It can be checked, see appendix \ref{appendixflip}, that
these terms appear in the expansion of the exponential in
(\ref{Aver}), as well as other terms that do not obey the central
charge restriction (\ref{ccrestriction}), but which will vanish when
projected on the spin states which do obey the restriction.  The
expression (\ref{Aver}) is reminiscent of a boundary state in
conformal field theory\footnote{The idea that the vertex should be
similar to a boundary state was suggested to us by R. Janik.}.

Let us now determine how the two versions of the vertex, $|V_{12}\rangle$ and $|\CV_{12}\rangle$ transform the oscillators from one space into the others.  ($ i=1,2$)
\begin{align}
  \label{exchangeosc}
(a^{(1)\dagger}_{i,s}+ b^{(2)}_{i,s}) |V_{12}\rangle
=(b^{(1)\dagger}_{i,s}- a^{(2)}_{i,s}) |V_{12}\rangle&=(a^{(1)}_{i,s}+
b^{(2)\dagger}_{i,s}) |V_{12}\rangle =(b^{(1)}_{i,s}-
a^{(2)\dagger}_{i,s}) |V_{12}\rangle =0\;, \nonumber\\
(c^{(1)}_{i,s}+ d^{(2)\dagger}_{i,s})
|V_{12}\rangle=(d^{(1)}_{i,s}+c^{(2)\dagger}_{i,s})
|V_{12}\rangle&=(d^{(1)\dag}_{i,s}-c^{(2)}_{i,s})
|V_{12}\rangle=(c^{(1)\dag}_{i,s}-d^{(2)}_{i,s}) |V_{12}\rangle=0 \; .
\end{align}
We have chosen the vertex (\ref{Aver}) $|V_{12}\rangle$ such as to
transform operators $(a_i,\,b_i ,\,c_i, \,d_{i})$ into
$(b_i^\dagger,\, a_i^\dagger, \,d_{i}^\dag, \,c_{i}^\dag)$, very much
as the action of the operator $\U^2$ in (\ref{ucarreosc}) does.  Let us look at the effect of the vertex on the generators of the
$\mathfrak{psu}(2,2|4)$ algebra.  In general, the vertex transforms
generators acting in one of the Fock spaces, $G^{(1)}$, into operators
acting in the other space, $\tilde G^{(2)}$, by
\begin{align}
  \label{vertexexchange}
  G^{(1)} |V_{12}\rangle &\equiv -\tilde G^{(2)} |V_{12}\rangle\;,
  \nonumber \\
G^{(1)}H^{(1)} |V_{12}\rangle &=(-1)^{|G||H|}\tilde H^{(2)} \tilde
G^{(2)} |V_{12}\rangle\;,
\end{align}
with $|G|$ denoting the grading of the operator $G$, {\it i.e.} the
number of fermions it contains modulo 2.  The transformation above is
an anti-morphism, because it changes the order of the operators.  Let
us consider the generators of the $\mathfrak{psu}(2,2|4)$ algebra (or
rather $\mathfrak{u}(2,2|4)$, since we prefer not to factor out the
central element and the super identity) $E^{AB(1)}$ which obey the
commutation relations (\ref{commrelsu}).  According to
(\ref{vertexexchange}), they are transformed by the vertex into
another set of generators, $\tilde E^{AB(2)}$, also obeying the
commutation relations\footnote{We have introduced the minus sign in
the first line of (\ref{vertexexchange}) to get the right commutation
relations for $\tilde E^{AB(2)}$ . } of $\mathfrak{psu}(2,2|4)$, and a
priori different from $E^{AB(2)}$.  We deduce that the vertex obeys
the local symmetry condition
%
\begin{align}
  \label{vertexsym}
\left( E_s^{AB(1)}+ \tilde E_s^{AB(2)}\right)
|V_{12}\rangle=0\;, \qquad s=1,\ldots L\;.
\end{align}

The explicit form of $\tilde E^{AB}$ can be determined using
(\ref{exchangeosc}) and (\ref{vertexexchange}).
We have, for example, for generators of the conformal subalgebra,
\begin{align}
  \label{scvertexsym}
\tilde L^+_\mu= -b \bar \sigma_\mu & a=U^2L^+_\mu U^{-2}\;, \qquad
\tilde L^-_\mu= b^\dagger \sigma_\mu a^\dagger=U^2L^-_\mu U^{-2}\;,
\nonumber\\
\tilde E&=-\frac{1}{2}(a a^\dagger + b^\dagger b)=U^2 E U^{-2}=-E\;.
\end{align}
By inspection, we can see that
\begin{align}
\tilde E^{AB}=\U^2 (E^{AB} +(-1)^{|B|}\delta^{AB})\U^{-2}
\end{align}
for all the generators, even and odd, with $|B|=0,1$ for bosonic and
fermionic indices respectively.  We therefore conclude that the
symmetry of the vertex $|\CV_{12}\rangle$, at tree level, can be expressed as
\begin{align}
  \label{vertexsymU}
\left( E_s^{AB(1)}+ E_s^{AB(2)}+(-1)^{|B|}\delta^{AB}\right)
|\CV_{12}\rangle=0\;, \qquad s=1,\ldots L\;.
\end{align}
The term $(-1)^{|B|}\delta^{AB}$ is proportional to the identity in the oscillator space and it can be incorporated into a shift of the Cartan generators, $E^{AA}\to E^{AA}+(-1)^{|A|}$, which does not affect the $\mathfrak{u}(2,2|4)$ commutation relations. Moreover, this shift preserves the central element
$\sum_AE^{AA}$;
we therefore conclude that the vertex possess local
$\mathfrak{psu}(2,2|4)$ symmetry.  Equation (\ref{vertexsymU})
justifies a posteriori the relation (\ref{symL}) we have used in the
definition of the correlation function.  This local symmetry can be
taken as a defining property of the vertex, and it will be deformed at
higher loop.

\subsection{Properties of the vertex}

In this section we are exploiting the properties of the vertex which
are useful for the computation of the correlation functions at the
tree level.  The first step is to characterize the states that are
flipped with the help of the vertex.  For this purpose, we work out
first the action of the monodromy matrix on the vertex and then
identify the flipped states.  The second step, which can be performed
in the $\mathfrak{so}(6)$ sector, is to separate the space-time
dependence from the structure constant and rederive the expression of
the structure constants in terms of the spin chain overlaps.  In
particular, in the $\mathfrak{so}(4)$ subsector we rederive the EGSV
\cite{EGSV} factorization of the structure constants.

\subsubsection{Characterizing the flipped operator $\bO$}
\label{subsec:barO}

\begin{figure}
         \centering
	 \begin{minipage}[t]{0.6\linewidth}
	    \centering
            \includegraphics[width=7.9 cm]{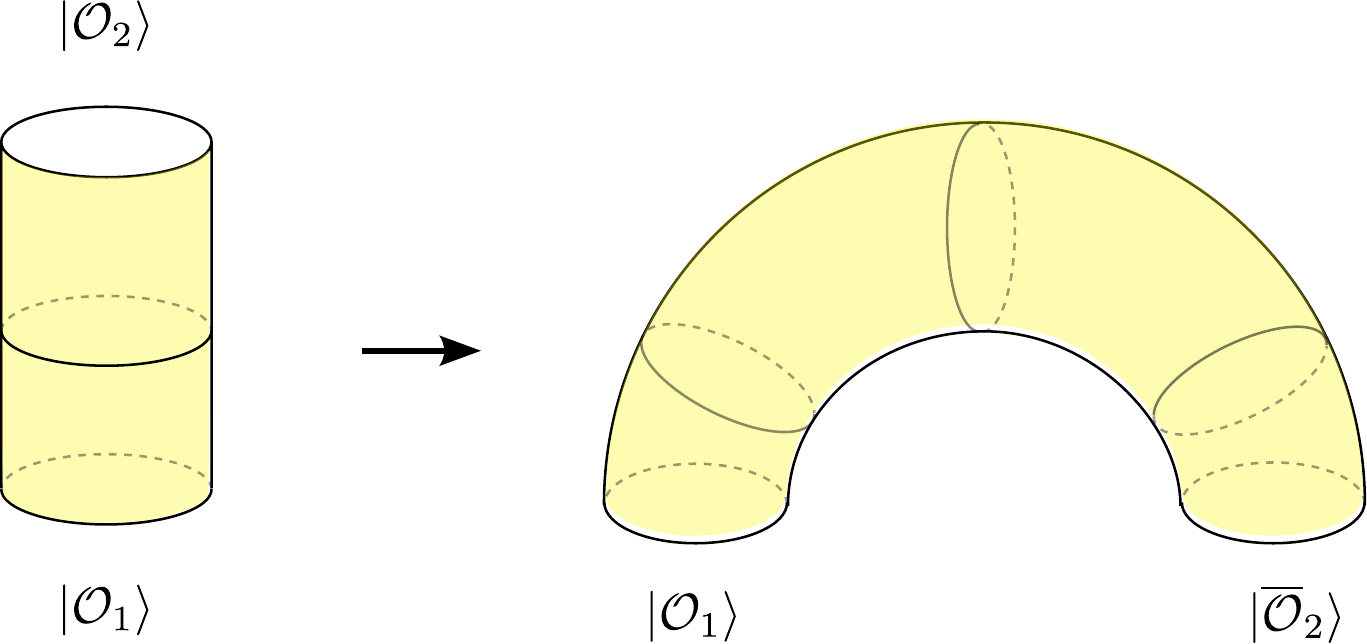}
  \caption{ Two ways of computing the two point function
and how to flip the operator $|\CO_2\rangle$ to $|\bO_2\rangle$. }
 \label{figsix}
         \end{minipage}
                   \end{figure}

One of the basic property of the vertex is that it transforms an
outgoing state into a incoming one (or vice versa),
\begin{align}
  \; ^{(1)}\langle{\cal O}|=\langle V_{12}| \sigma^{(1)}|\bO
  \rangle^{(2)}\;,
\end{align}
the two states $\langle{\cal O}|$ and $| \bO \rangle$ corresponding to
two different but related operators $\CO$ and $\bO$.  In this section,
we are going to show how to obtain the operator $\bO$ once $\CO$ is
given.  In this way we are relating the two different way of computing
the two point functions illustrated in figure \ref{figsix}.

Due to the large degeneracy of trace states at tree level, one prefers
to use a pre-diagonalization and use as basis of states the
eigenstates of the one-loop dilatation operator, which is conveniently
given by (nested) algebraic Bethe ansatz.  Suppose that we have built
the one-loop Lax matrix
\begin{align}
\label{laxdef}
L_s^{}(u)=u-i/2-i(-1)^{|A|}E_{0}^{AB}E_s^{BA}\;.
\end{align}
where the generators in the auxiliary space $E_{0,d}^{AB}$ belong to
the defining $(4|4$ dimensional) representation of
$\mathfrak{psu}(2,2|4)$ and $E^{AB}_s$ are the generators in the
actual physical representation, {\it e.g.} the oscillators
representation.  Using the property (\ref{vertexsymU}) of the vertex
it is straightforward to show that
\begin{align}
\label{L12}
L^{(1)}(u) |\CV_{12}\rangle =-L^{(2)}(-u) |\CV_{12}\rangle\;.
\end{align}
Since the vertex carries the physical representation and its dual, one
could interpret the above relation as the crossing relation.  This
point can be made more explicit by using the set of generators $\bar
E^{AB}$ defined in (\ref{conjgen}) which act naturally in the dual
representation.  The change of sign in the Lax matrix can be absorbed
in the normalization, and we will tacitly assume in the following that
we have done so.  Let us now consider the monodromy matrices of the
two chains
\begin{align}
 T^{(1)}(u)=L_1^{(1)}(u)\ldots L_L^{(1)}(u)\;, \qquad
 T^{(2)}(u)=L_L^{(2)}(u)\ldots L_1^{(2)}(u)
\end{align}
 and apply repeatedly the relation (\ref{L12}).  We remind the
 convention (\ref{revord}) for the order of the sites of the second
 chain.  The result is
\begin{align}
\label{Tvertex}
L_1^{(1)}(u)\ldots L_L^{(1)}(u)|\CV_{12}\rangle =L_1^{(2)}(-u)\ldots
L_L^{(2)}(-u)|\CV_{12}\rangle\;.
\end{align}
The right hand side is not exactly the monodromy matrix for the second
chain $T^{(2)}(u)$, because the Lax matrices are in reverse order.
This mismatch can be cured by taking an operation which reverses the
order of the operators, like the (super) transposition $^{t_0}$ in the
auxiliary space.
In some sectors of $\mathfrak{psu}(2,2|4)$ one can correlate the
change of the signs of the supertraceless generators $E^{ab}$ with the
transposition
\begin{align}
E^{ab}=-\sigma E^{ab,t}\sigma^{-1}\;.
\end{align}
where $^t$ denotes the (super) transposition in the quantum space.
This is the case, for example, for the $\mathfrak{so}(4)\simeq
\mathfrak{su}(2)_L\otimes \mathfrak{su}(2)_R$ sector, where
$\sigma=\sigma^{-1}=-\sigma_{2,L}\sigma_{2,R}=U_F^{2}$.  As one can
check on (\ref{laxdef}), in any of the $ \mathfrak{su}(2)$ sectors we
have
\begin{align}
L(u)=L^{t_0,t}(u)=-\sigma L^{t_0}(-u)\sigma^{-1}=-\sigma_0
L^{t_0}(-u)\sigma_0^{-1}\;,
\end{align}
where $\sigma_0=i\sigma_{2,0}$.  The last equality sign comes from the
invariance of the Lax matrix $[L_s(u), E^{ab}_0+E^{ab}_s]=0$.
Substituting one of the last two equalities above into the r.h.s. of
in (\ref{Tvertex}) we obtain\footnote{We neglect again an overall
normalization.}
\begin{align}
T^{(1)}(u)|\CV_{12}\rangle=\sigma\,
T^{(2),t_0}(u)\sigma^{-1}|\CV_{12}\rangle=\sigma_0
T^{(2),t_0}(u)\sigma_0^{-1}|\CV_{12}\rangle\;.
\end{align}
or in matrix form
\begin{align}
\label{vertbop}
 \left(\begin{array}{cc}A(u) & B(u)\\C(u) &
 D(u)\end{array}\right)^{\!\!\!  (1)}\!\!\!
 |\CV_{12}\rangle&=\left(\begin{array}{cc}\sigma A(u)\sigma ^{-1} &
 \sigma C(u)\sigma ^{-1} \\\sigma B(u)\sigma ^{-1} & \sigma D(u)\sigma
 ^{-1}\end{array}\right)^{\!\!\!
 (2)}|\CV_{12}\rangle=\left(\begin{array}{cc}D(u) & -B(u)\\-C(u) &
 A(u)\end{array}\right)^{\!\!\!  (2)}\!\!\!  |\CV_{12}\rangle\;.
\end{align}
We will exemplify now the consequence of these relation in a given
$\mathfrak{su}(2)$ sub-sector.  The eigenvectors of the dilatation
operator can be constructed by the action of the $B$ operators on the
vacuum state $|Z^L\rangle$ followed by an arbitrary $\mathfrak{so}(6)$
rotation $\R$ in the quantum space,
\begin{align}
| \CO \rangle=\R \; { B}(u_1)\ldots { B}(u_M)|Z^L\rangle\;.
\end{align}
The global rotation $\R$ changes the orientation of the
$\mathfrak{su}(2)$ sector inside $\mathfrak{so}(6)$.  Let us note that
if we descend to $\mathfrak{so}(4)$, there are two different orbits
the $\mathfrak{su}(2)$ sectors inside $\mathfrak{so}(4)$, called in
the literature $\mathfrak{su}(2)_R$ and $\mathfrak{su}(2)_L$ and
obtained by rotating $(Z,X)$ and $(Z,\bar X$) respectively.  The two
orbits are related to each other by improper rotations.
Since we are working with operators which do not have components
outside the $\mathfrak{so}(6)$ sector, we are going to use a version
of the vertex $\langle v_{12}|$ truncated to $\mathfrak{so}(6)$.  By
equation (\ref{vertbop}) we obtain the rule which transfers the Bethe
operators from one space to the other through the vertex, \footnote{A
similar relation was known to S. Komatsu \cite{ShotaPohang}.}
\begin{align}
\nonumber \langle v_{12}|[\R \,&B(u_1)\ldots B(u_M)]^{(1)}=\langle
v_{12}|[B(u_M)\ldots B(u_1)\,\R^{-1}]^{(2)}\\
 &=\langle v_{12}|[\sigma \,C(u_M)\ldots C(u_1)\sigma \R^{-1}
 ]^{(2)}\;.
\end{align}
This relation is fundamental in exploiting the vertex, and it
prescribes in particular how to characterize the flipped states
\begin{align}
&\; ^{(2)}\langle{\bO}|= \langle v_{12}|\CO\rangle^{(1)}=\\ =\langle
v_{12}|\;[B(u_M)\ldots
B(u_1)&\,\R^{-1}]^{(2)}\;|Z^L\rangle^{(1)}=\,^{(2)}\langle \bar
Z^L|\;[B(u_M)\ldots B(u_1)\,\R^{-1}]^{(2)}\nonumber\\\nonumber
=\langle v_{12}|\;[\sigma\, C(u_M)\ldots &C(u_1)\,\sigma
\,\R^{-1}]^{(2)}\;|Z^L\rangle^{(1)}=\,^{(2)}\langle
Z^L|\;[C(u_M)\ldots C(u_1)\,\sigma\, \R^{-1}]^{(2)}
\end{align}
 Using $B(u)^\dagger=-C(u^*)$ and considering distributions of
 rapidities which are self-conjugate, $\{u\}=\{u^*\}$ we conclude
 that, up to an overall sign,
\begin{align}
\label{bO}
| \bO \rangle=\R\, C(u_1)\ldots C(u_M)\,|\bar
Z^L\rangle=\R\,\sigma\,B(u_1)\ldots B(u_M)\,| Z^L\rangle\;.
\end{align}
 Keeping in mind that $| \bO \rangle$ lives in a spin chain with the
 order of the site reversed with respect to $| \CO \rangle$ we
 conclude that this is essentially the flipping procedure of
 \cite{EGSV}.  The alternative definitions of the Bethe vectors like
 in (\ref{bO}) can be used at will in order to express the overlaps in
 a convenient form.  For example the last equality in the above
 equation can be proven to be equivalent to the result by one of the
 authors and Y. Matsuo \cite{SZ} that the scalar product of one
 on-shell and one off-shell Bethe state are Izergin determinants.

\subsubsection{Tree level correlation function in the
$\mathfrak{so}(6)$ sector and the overlaps}
\label{subsection:EGSV}

As we have already seen in equation (\ref{corrtree}), the two point
function at tree level in the $\mathfrak{so}(6)$ sector can be reduced
to the computation of an overlap,
\begin{align}
 \label{corrtreeop}
\langle \CO_1 (x) \CO_2(y) \rangle = \langle \bO_1|U^2 e^{iL^+(y- x)}
|\CO_2\rangle =\frac{ \langle
\bO_1|\CO_2\rangle}{(x-y)^{2\Delta_1}}=\frac{\langle v_{12}|
\CO_1\rangle\otimes|\CO_2\rangle}{(x-y)^{2\Delta_1}}\;,
\end{align}
where again $\langle v_{12}|$ is the vertex $\langle \CV_{12}|$
reduced to the $\mathfrak{so}(6)$ sector.  The same is valid for the
three point function at tree level,
\begin{align}
\nonumber \langle {\cal O}_2(x_2)&  {\cal O}_3(x_3) {\cal O}_1(x_1)
\rangle= \\
  &= \langle V_{123}|\, \U_{(13)}^2\, \U_{(12)}^2 \U_{(32)}^2 \;
  e^{i[ L^+_{(1)}x_1+ L^+_{(2)} x_2+ L^+_{(3)}x_3]} |\CO_2\rangle
  \otimes | \CO_3\rangle \otimes |{\cal O}_1\rangle \nonumber
\end{align}
\vskip-15pt
\begin{align}
 \label{mapttpfverop}
&=\frac{\langle v_{123}|\CO_2\rangle \otimes | \CO_3\rangle \otimes
|{\cal
O}_1\rangle}{|x_{12}|^{\Delta_{12}}|x_{13}|^{\Delta_{13}}
|x_{23}|^{\Delta_{23}}}\,,
\end{align}
\vskip3pt \noindent where $\Delta_{ij}=\Delta_i+\Delta_j-\Delta_k$
with $\{i,j,k\}=\{1,2,3\}$.  To obtain this relation we use that at
tree order we can freely split the chain $(i)$ into two pieces $(ij)$
and $(ik)$ which connect with chains $(j)$ and $(k)$ respectively, and
\begin{align}
L_{(i)}^+=L_{(ij)}^++L_{(ik)}^+\;, \qquad \langle \CV_{123}|[
L_{(ij)}^++L_{(ji)}^+]=0\;,
\end{align}
then we use the normal form (\ref{norf}) of the operators $U_{(ij)}^2$
to evaluate the averages over the bosonic oscillators.  The separation
of space-time dependence and the structure constant is possible in the
sectors that do not contain bosonic oscillators.  In sectors which
contain bosonic oscillators, like $\mathfrak{sl}(2)$ and
$\mathfrak{su}(1|1)$, one can have typically several tensor structures
for the space-time dependence \cite{2011arXiv1109.6321C,
Costa:2011mg}.  So, in the $\mathfrak{so}(6)$ sector we can reduce the
structure constant to the overlap
\begin{align}
C_{123}=\langle v_{123}|\CO_2\rangle \otimes | \CO_3\rangle \otimes
|{\cal O}_1\rangle\;,
\end{align}
where we suppose that the states $|\CO_i\rangle$ are normalized,
${\cal N}_i= \langle \CO_i |\CO_i\rangle=1$.  If this is not the case,
one has to divide out $\sqrt{{\cal N}_1{\cal N}_2{\cal N}_3}$.

We would like now to discuss more in detail the correlation functions
of three operators in different $\mathfrak{su}(2)$ sectors, since they
have been studied in detail in the literature
\cite{2012arXiv1212.6563K,2011JHEP09132G,Sobko:SoV}.  As we have
already mentioned, there are two different orbits of the
$\mathfrak{su}(2)$ sectors under the global $\mathfrak{so}(4)$
rotations, and we will call them after the $\mathfrak{su}(2)_R$ and
$\mathfrak{su}(2)_L$ defined below.  We take the convention
\begin{align}
|Z\rangle =|0\rangle\;, \qquad |\bar Z\rangle =c_1^\dag d_1^\dag
c_2^\dag d_2^\dag|0\rangle\;, \qquad |X\rangle =c_1^\dag
d_1^\dag|0\rangle\;,\qquad |\bar X\rangle =-c_2^\dag
d_2^\dag|0\rangle\;.
\end{align}
and that the $L$ sector is generated by $c_1,d_1$ and the $R$ sector
by $c_2,d_2$.  Obviously, the generators in the two sectors commute,
and the operators $X, \bar X, Z, \bar Z$ can be seen as basis vectors
in the bi-fundamental representation of $\mathfrak{su}(2)_R \otimes
\mathfrak{su}(2)_L$,
\begin{align}
&|Z\rangle =|\!\uparrow\rangle_L\otimes | \!\uparrow
\rangle_R\equiv|\!\uparrow\uparrow\rangle\;, \qquad |\bar Z\rangle
=|\!\downarrow\rangle_L\otimes |
\!\downarrow\rangle_R\equiv|\!\downarrow\downarrow\rangle\;,\\
& |X\rangle = |\!\uparrow \rangle_L\otimes |\!
\downarrow\rangle_R\equiv|\!\uparrow\downarrow\rangle\;,\qquad |\bar
X\rangle =-|\!\downarrow\rangle_L\otimes |
\!\uparrow\rangle_R\equiv-|\!\downarrow\uparrow\rangle\;.  \nonumber
\end{align}
The authors of \cite{ShotaNew} call this representation the double
spin, or double chain, representation, which can be traced back to
\cite{Kazakov:2004qf}.  Together, the two $\mathfrak{su}(2)$ sectors
generate an $\mathfrak{so}(4)$ sector.  The vertex reduced to this
sector is
\begin{align}
 \nonumber |v_{12}\rangle^{\mathfrak{so}(4)} =&|Z\rangle\otimes |\bar
 Z\rangle +|X\rangle\otimes |\bar X\rangle +|\bar Z\rangle\otimes
 |Z\rangle+|\bar X\rangle\otimes
 |X\rangle=|v_{12}\rangle^{\mathfrak{su}(2)_L}\otimes
 |v_{12}\rangle^{\mathfrak{su}(2)_R}\;,\\
&|v_{12}\rangle^{\mathfrak{su}(2)_{L,R}}=|\!\uparrow\rangle_{L,R}
\otimes|\!\downarrow\rangle_{L,R}-|\!\downarrow\rangle_{L,R}\otimes
|\!\uparrow\rangle_{L,R}\;.
\end{align}
We can have two different cases:

$i)$ The $RRR$ case, when all the three operators are in the same
sector, say $R$.  In this case, the three operators can be chosen as
\begin{align}
 |{\cal O}_1\rangle&= \R_1 B_R(u_1)\ldots B_R(u_{M_1})\,
 |Z^{L_1}\rangle\;,\\
 |{\cal O}_2\rangle&= \R_2\, \sigma \,B_R(v_1)\ldots B_R(v_{M_2})\,
 |Z^{L_2}\rangle\;,\nonumber \\
 |{\cal O}_3\rangle&= \R_3\, \sigma \, B_R(w_1)\ldots B_R(w_{M_3})\,
 |Z^{L_3}\rangle\;.\nonumber
\end{align}
The convention is such that $\R_1=\R_2=\R_3=1$ reduces to the extremal
case.\footnote{In the extremal case one has to take into account the
effect of mixing with higher trace operators, which is not done here.
We thank S. Komatsu for mentioning to us that there exist non-extremal
$RRR$ correlators.  } Although the explicit computation of the
structure constants goes beyond the scope of this paper, we can note
that this case does not seem to be computable in the generic case
without cutting the states into pieces as prescribed by \cite{EGSV}.

$ii)$ The $RRL$ case, when two operators, say $\CO_1$ and $\CO_2$, are
in the sector $R$ and $\CO_3$ is in the sector $L$.  In this case we
choose
\begin{align}
 |{\cal O}_1\rangle&= \R_1 B_R(u_1)\ldots B_R(u_{M_1})\,
 |Z^{L_1}\rangle\;,\\
 |{\cal O}_2\rangle&= \R_2\, \sigma\, B_R(v_1)\ldots B_R(v_{M_2})\,
 |Z^{L_2}\rangle\;,\nonumber \\
 |{\cal O}_3\rangle&= \R_3\, B_L(w_1)\ldots B_L(w_{M_3})\,
 |Z^{L_3}\rangle\;.\nonumber
\end{align}
Again, our choice is such that $\R_1=\R_2=\R_3=1$ is the case
originally considered by EGSV \cite{EGSV}.  In this case the left and
right sector decouple
\begin{align}
C^{EGSV}_{123}=\,^{\mathfrak{so}(4)}&\langle v_{123}|\
B_L(w)\,|Z^{L_3}\rangle\; \otimes\; \sigma_{(2)}\, B_R(v)\,|
Z^{L_2}\rangle \; \otimes \; B_R(u)\,| Z^{L_1}\rangle\\\nonumber
=\,^{\mathfrak{so}(4)}\langle v_{123}&|\
\sigma_{(32)}\,B_L(w)\,|Z^{L_3}\rangle \; \otimes \; \sigma_{(21)}\,
B_R(v)\,| Z^{L_2}\rangle \; \otimes \; B_R(u)\,| Z^{L_1}\rangle\;.\\
&={\rm SIMPLE}\times {\rm INVOLVED}
\end{align}
The {\rm SIMPLE} part is given by the contribution of the $L$ sector,
\begin{align}
{\rm SIMPLE}=\,^{\mathfrak{su}(2)_L}\langle v_{123}|\
\sigma_{(32)L}\,B_L(w)\,|\uparrow^{L_3}\rangle\,\otimes &
\;\sigma_{(21)L}\,|\uparrow^{L_2}\rangle\otimes|\uparrow^{L_1}\rangle
\nonumber\\
=\langle
\downarrow^{L_3}|\,\sigma_{(32)L}\,B_L(w)\,|\uparrow^{L_3}\rangle &
\end{align}
while {\rm INVOLVED} is given by the contribution of the $R$ sector
\begin{align}
{\rm INVOLVED}&=\,^{\mathfrak{su}(2)_R}\langle
v_{123}|\;\sigma_{(32)R}\,|\uparrow^{L_3}\rangle\otimes
\,\sigma_{(21)R}\,B_R(v)\,|\uparrow^{L_2}\rangle\otimes
B_R(u)\,|\uparrow^{L_1}\rangle\nonumber\\
&=\,^{\mathfrak{su}(2)_R}\langle v_{12}|\;\sigma_{(21)R}\ \langle\,
\uparrow^{L_{23}}|\,B_R(v)\,|\uparrow^{L_2}\rangle \otimes \langle\,
\downarrow^{L_{13}}|B_R(u)\,|\uparrow^{L_1}\rangle
\end{align}
Now one can use the properties of the Bethe states to show that
\begin{align}
\langle\,\uparrow^{L_{23}}|\,B_R(v)\,|\,\uparrow^{L_2}\rangle=B_R(v)\,
|\uparrow^{L_{21}}\rangle=\langle\,\downarrow^{L_{13}}|\,
B_R(i/2)^{L_{13}}\,B_R(v)\,|\uparrow^{L_1}\rangle\;,
\end{align}
where the $L_{13}$ operators $B(i/2)$ are freezing $L_{13}$
consecutive sites to their $\downarrow$ value \cite{Omar}.  This
implies also that freezing selects a single component from the vertex
$\langle v_{13}|$
\begin{align}
\,^{\mathfrak{su}(2)_R}\,\langle v_{13}|\,\sigma_{(13)R}
\,B_R(i/2)^{L_{13}}\,B_R(v)\,|\uparrow^{L_1}\rangle=\langle\,
\downarrow^{L_{13}}|\otimes \langle\,\downarrow^{L_{13}}|\,
B_R(i/2)^{L_{13}}B_R(v)|\uparrow^{L_1}\rangle
\end{align}
we obtain finally
\begin{align}
{\rm INVOLVED}&=\,^{\mathfrak{su}(2)_R}\langle v_{12}|\,\langle
v_{13}|\,\sigma_{(1)R}\,B_R(i/2)^{L_{13}}
B_R(v)\,|\uparrow^{L_{1}}\rangle \otimes
B_R(u)\,|\uparrow^{L_{1}}\rangle\;.
\end{align}
So we have transformed the involved part into an overlap involving a
single spin chain of length $L_1$.  This is the result of EGSV
\cite{EGSV} combined with O. Foda's freezing trick \cite{Omar}.  The
case when the global rotations $\R_1,\ \R_2,\ \R_3$ are arbitrary is
considered in \cite{ShotaNew}.

\subsection{Scalar products and global $\mathfrak{su}(2)$ rotations}

Although considering correlators with the global $\mathfrak{so}(6)$
rotations goes beyond the scope of this work, it is relatively simple
and instructive to consider the scalar product of two
$\mathfrak{su}(2)$ Bethe states $|\{u\}\rangle$ and $|\{v\}\rangle$
that are rotated with respect to each other with an $\mathfrak{su}(2)$
rotation,
\begin{eqnarray}
\R_{\mathfrak{su}(2)}= e^{a \sigma^+}e^{ia_3\sigma_{3}}e^{-\bar
a\sigma^-}\;.
\end{eqnarray}
By expanding the left and right factors in the rotation, $e^{a
\sigma^+}=\sum_{k\geq 0} a^k (\sigma^+)^k/k!$ and supposing that
$\{u\}$ and $\{v\}$ contain the same number of mangons $M$, we get
\begin{eqnarray}
\label{scprodR}
\langle\{u\}|\R_{\mathfrak{su}(2)}|\{v\}\rangle=e^{ i a_3
(L-2M)}\sum_{k= 0}^{L-2M}\frac{(-a\bar a e^{-2 i a_3 })^k}{(k!)^2}
\langle\{u\};k|k;\{v\}\rangle\;,
\end{eqnarray}
with the state $|k;\{v\}\rangle$ containing $k$ magnons at infinity.
The reason that the sum stops at $L-2M$, and not at $L-M$, as one
could naively think, is that the state $|\{v\}\rangle$ is the highest
weight state of a multiplet with spin $L/2-M$ and as such one cannot
act on it more than $L-2M$ times with lowering operators.  As shown in
appendix \ref{apps:sep}, if at least one of the states $|\{u\}\rangle$
and $|\{v\}\rangle$ is on-shell, the scalar products with $k$ magnons
sent to infinity is given by
\begin{eqnarray}
\label{magntriv}
\langle\{u\};k|k;\{v\}\rangle=(k!)^2 {L-2M \choose k}
\langle\{u\}|\{v\}\rangle\;.
\end{eqnarray}
After resumming the sum in (\ref{scprodR}) one obtains the simple
expression
\begin{eqnarray}
\langle\{u\}|\R_{\mathfrak{su}(2)}|\{v\}\rangle=(e^{ i a_3 }-a \bar a
e^{-ia_3})^{L-2M} \langle\{u\}|\{v\}\rangle\;.
\end{eqnarray}
It is interesting and reassuring to note that this relation holds when
the scalar product $\langle\{u\}|\{v\}\rangle$ can be put in a
determinant expression.  It would be interesting to check whether this
relation hold for more general rotations, for example in
$\mathfrak{su}(3)$, where determinant expressions for states with some
set of magnons at infinity also exist \cite{Wheeler-SU3}.

\section{Monodromy condition on the spin vertex}
\label{section:Mono}

In this section we are going to show that the local symmetry condition
(\ref{vertexsymU}) of the spin vertex can be reformulated as an
extended symmetry. This is the same  Yangian symmetry, satisfied
by the tree-level amplitudes in ${\cal N}=4$ SYM \cite{Drummond:Yangian}.

The spin vertex is an invariant of the Yangian.  We are going first to
show this on the two-vertex, and then extend it to the three-vertex we
need to compute the three point function.  There are two types of
monodromy matrices which are interesting for us.  The first is the
monodromy matrix where the auxiliary space is in the defining, $4|4$
dimensional, representation.  This monodromy matrix is useful to build
the Yangian generators and the for the nested Bethe ansatz procedure.
The second type of monodromy matrix, useful for getting the local
conserved quantities, contains the same physical representation in the
auxiliary and quantum spaces.  Here we construct the monodromy matrix
with the auxiliary space in the defining representation.  For the
monodromy matrix with the auxiliary space in the physical
representation, the construction of the $\mathfrak{so}(6)$ sector is
relatively straightforward, however the construction in the
$\mathfrak{sl}(2)$ sector is more subtle and we are not doing it here.

\begin{figure}
         \centering
	 \begin{minipage}[t]{0.7\linewidth}
            \centering
            \includegraphics[width=10.9 cm]{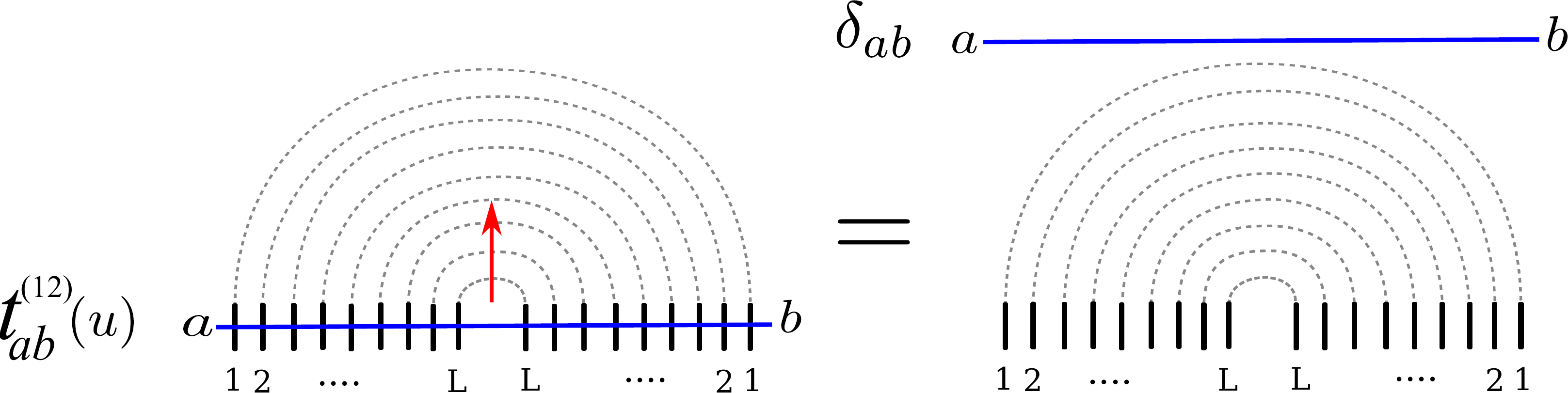}
  \caption{ The two chain monodromy matrix $t^{(12)}(u)$
 and its action on the vertex $|\CV_{12}\rangle$}
 \label{figfive}
         \end{minipage}
         \bigskip
                   \end{figure}

Let us take the $\mathfrak{psu}(2,2|4)$ $R$ matrix in the defining and
physical representation
\begin{align}
\label{Rdef}
R_{01}(u)=u-i\Pi_{01}\;, \qquad
\Pi_{01}=(-1)^{|A|}E_{0}^{AB}E_1^{BA}\;,
\end{align}
where $E_{0}^{AB}$ are $4|4\times 4|4$ super matrices and the
generators in the quantum space are in the oscillator representation
$E_1^{BA}=\bar \psi^{A}\psi^B$.  When $E_1^{BA}$ are also in the
defining representation, $\Pi_{01}$ is a super-permutation.  In the
representation we are considering
\begin{align}
\label{Pisq}
\Pi_{01}^2&=(-1)^{|A|+|C|}E_{0}^{AB}E_1^{BA}E_{0}^{CD}E_1^{DC}
=(-1)^{|A|+|B|+(|A|+|B|)(|B|+|D|)}E_0^{AD}E_1^{BA}E_1^{DB}\\
\nonumber
&=(-1)^{|A|}E_0^{AD}E_1^{DA}(E_1^{BB}-1)+E_1^{BB}
=\Pi_{01}(E_1^{BB}-1)+E_1^{BB}=-\Pi_{01}\;.
\end{align}
Here we have used the (anti)commutation relations $[\psi^A,\bar \psi^{
B}]_\pm=\delta^{AB}$ and that in the physical representation
$c=E^{BB}_1=\bar \psi^{ B}\psi^B=0$ \footnote{ The condition $c=0$
should be understood as a constraint imposed on the states, which
projects on the irreducible representation we are interested in.  This
constraint can be implemented in the definition of the spin vertex,
but then the vertex will lose its nice exponential form.} and in the
auxiliary representation $E_0^{BB}=1$.  The $R$ matrix above satisfies
the unitarity condition
\begin{align}
\label{Runit}
R_{01}(u)R_{01}(-i-u)=-u(i+u)\;.
\end{align}
For a representation with arbitrary central charge $c$, the unitarity
condition would be
\begin{align}
\label{Runitc}
R_{01}(u)R_{01}(i(c-1)-u)=-u(i(1-c)+u)-c\;.
\end{align}

We are now going to build the monodromy condition for the two-site
vertex $|\CV_{12}\rangle$,
\begin{align}
\label{moncd}
R_{01} (u)R _{02}(u)|\CV_{12}\rangle=-R_{01} (u)R_{01}
(-i-u)|\CV_{12}\rangle=u(u+i)|\CV_{12}\rangle\;.
\end{align}
Here we have used that the $R$ matrix is related to the Lax matrix
defined in (\ref{laxdef}) by $R_{01}^{(2)}(u)=L^{(2)}(u+i/2)$, and
then use the crossing-like property (\ref{L12}) of the vertex
\begin{align}
\label{cross}
R_{02}(u)|\CV_{12}\rangle=-R_{01}(-i-u)|\CV_{12}\rangle.
\end{align}
The condition (\ref{moncd}) can be lifted to the two-vertex with an
arbitrary number of sites, as depicted in figure \ref{figfive}
\begin{align}
\label{monmatver}
T_{12}(u)=R^{(1)}_{01}(u)\ldots R^{(1)}_{0L}(u)\;
R^{(2)}_{0L}(u)\ldots R^{(2)}_{01}(u)\;
|\CV_{12}\rangle=(u(u+i))^L|\CV_{12}\rangle\;.
\end{align}
as well as for the three vertex, where the different pieces
$t^{(ij)}(u)$ joining chain $(i)$ to chain $(j)$ are glued as in
figure \ref{3Vshift},
\begin{align}
\label{tpvmonobis}
T_{123}(u)= t^{(12)}(u) t^{(13)}(u) t^{(31)}(u) t^{(32)}(u)
t^{(23)}(u) t^{(21)}(u)\;.
\end{align}

\noindent {\bf The subsectors:} \vskip5pt \noindent The
$\mathfrak{psu}(2,2|4)$ $R$ matrix can be readily reduced to different
subsectors, just by restricting the sum in the definition of the
central charge \re{ccrestriction} to the corresponding subsector.  As
a result, the central charge can take non-zero value $c= E^{BB}_1$.

\begin{itemize}
\item
In the $\mathfrak{su}(1|1)$, $\mathfrak{su}(2|3)$ and
$\mathfrak{su}(2)$ sector, where the fields belong to the fundamental
representation, $c=1$, so that the unitarity condition is slightly
modified,
\begin{align}
\Pi_{01}^2=1\;, \qquad R_{01}(u)R_{01}(-u)=-(u^2+1)\;.
\end{align}
The monodromy condition will be
\begin{align}
\label{moncdbis}
R_{01}(u)R_{02}(u-i)|\CV_{12}\rangle
=-R_{01}(u)R_{01}(-u)|\CV_{12}\rangle=(u^2+1)|\CV_{12}\rangle\;.
\end{align}
\item
In the $\mathfrak{sl}(2)$ sector, $c=0$, so the unitarity and
monodromy conditions are the same as for $\mathfrak{psu}(2,2|4)$.
\item
In the $\mathfrak{so}(6)$ sector we have $c=2$, so that
\begin{align}
\Pi_{01}^2=\Pi_{01}+2\;, \qquad R_{01}(u)R_{01}(i-u)=u(i-u)-2\;.
\end{align}
The monodromy condition is then
\begin{align}
\label{moncdbisbis}
R_{01}(u)R_{02}(u-2i)|\CV_{12}\rangle=
-R_{01}(u)R_{01}(i-u)|\CV_{12}\rangle=(u(u-i)+2)|\CV_{12}\rangle\;.
\end{align}

\end{itemize}

\section{Conclusion and Outlook}
In this paper we proposed a new formulation for computing correlation
functions in planar $\mathcal{N}=4$ SYM theory.  In this novel
formalism, the central object is called the spin vertex, which is the
weak-coupling counter-part of the string vertex in the string field
theory.  We constructed the spin vertex for all sectors of the theory
at tree-level by a set of bosonic and fermionic oscillators.  The spin
vertex is a special entangled state living in Hilbert space of multi
spin chains and has many nice properties.  In the spin vertex
formalism, the symmetry of correlation functions become manifest.  In
particular, we are able to construct monodromy matrices under the
action of which the spin vertex is invariant.  In another word, the
spin vertex is invariant under the action of the infinite dimensional
Yangian algebra, which is the hallmark of integrability.\par

The spin vertex and its Yangian invariance is not only important
conceptually, but is also very useful practically.  Using the
properties of spin vertex in an ingenious way, the authors of
\cite{Shota1:Yangian} were able to compute more general configurations
of three-point functions both in the compact $SU(2)$ as well as in the
non-compact $SL(2)$ \cite{Shota2} sectors in terms of determinants.
In the semiclassical limit, the Yangian invariance of spin vertex is
equivalent to the monodromy condition which plays an important role in
the computation of three-points in the strong coupling limit
\cite{Komatsu:3pt1,Komatsu:3pt2,Komatsu:su2}.  This opens a new way of
computing semi-classical three-point functions by similar techniques
from strong coupling without using determinant formulas
\cite{ShotaNew}.\par

There are many open questions.  First and foremost, the present work
is inspired by the structure of the light-cone string field theory for
strings moving on the pp-wave background.  A natural question is
whether we can recover the light-cone string field theory in the BMN
limit.  The BMN limit is a degenerate limit of AdS/CFT correspondence
where all scattering phases are zero and hence integrability becomes
trivial.  However, it is interesting at both strong and weak coupling
to see how this limit is achieved.  This will be helpful to understand
the BMN limit better and might shed some light on finite coupling
regime.  At the leading order, we can show that the spin vertex in the
BMN limit reproduces exactly the structure of light-cone string field
theory with the same Neumann coefficients.  The derivation uses a
polynomial representation of the spin vertex and the result will be
presented elsewhere \cite{Jiang:Vertex}.

Another important question is understanding how to deform the spin
vertex and the corresponding Yangian invariance at higher loop orders
in perturbation theory.  In the computation of structure constants at
loop orders, quantum corrections manifest themselves as operator
insertions at the splitting points
\cite{GV:quantumintegrability,GV,JKLS-fixing,Pedro:sl23pt,Caetano:Fermionic}.
At present, these operator insertions are computed by Feynmann
diagrams which are usually rather complicated.  The generalization for
larger sectors and to higher loops in this way will be impractical.
However, since the theory is integrable, it should be possible to fix
these insertions from integrability, as in the case of the spectral
problem.  The higher loop deformation of the spin vertex should
contain the operator insertions at higher loops.  This problem is more
subtle due to renormalization.  In contrast to the tree-level, it is a
non-trivial task to extract the renormalization scheme independent
structure constant from the three-point function.  However, we think
that some general principles can still be applied.  We expect that at
higher loop the expression of the three point function is still given
by
\begin{align}
  \label{tpfhl}
\langle {\cal O}_2(y)  {\cal O}_2(y) {\cal O}_1(x)  \rangle = \langle
\CV_{123}|\, e^{i( L^+_1x+ L^+_2 y+ L^+_3z)} |\CO_2\rangle \otimes |
\CO_3\rangle \otimes |{\cal O}_1\rangle \,,
\end{align}
with all the quantities receiving radiative corrections.  The
space-time dependence of the correlator can be fixed by using Ward
identities, that can be derived for example by inserting the energy
operator $E_1+E_2+E_3$.  The constraints that the vertex has to
satisfy at any loop order is
\begin{align}
  \label{Wardvertex}
(E_1+E_2+E_3)| \CV_{123}\rangle=0 \,.
\end{align}
A similar constraint can be derived from the monodromy relation
(\ref{tpvmonobis}).  This suggests that the infinite Yangian symmetry
could be translated into Ward identities which would determine the
three-point correlation function.  We hope to be able to report on
this in the near future.\par

Finally, we would like to point out the similarity between our
construction of the spin vertex and the scattering amplitudes.
Yangian invariants were recently exploited to build the scattering
amplitudes
\cite{Ferro:2012xw,Chicherin-Kirchner,Chicherin,Frassek2013,
Florian-ampl,Staudacher:Yangian,Kanning:2014maa}.  Their key point is
to regard the scattering amplitudes as Yangian invariants and try to
construct it explicitly from Bethe ansatz.  To certain extent, the
spin vertex constructed in this paper is the simplest possible Yangian
invariant one can construct.  It is interesting to understand whether
more general Yangian invariants will play some role in the
construction of spin vertex, especially at higher loops.  In both
cases, the understanding of how to deform Yangian invariants at higher
loops is crucial.  This observation shows that Yangian invariant may
be the key to understand both on-shell quantities like scattering
amplitudes and off-shell quantities like correlation functions.  It
will be fascinating to develop a common framework and have a unified
description of these two kinds of quantities.

\acknowledgments

It is our pleasure to thank Z. Bajnok, R. Janik, N. Kanning, G.
Korchemsky, Y. Matsuo and especially S. Komatsu for very valuable
discussions.  Part of this work has been done during the visit of I.K.
and D.S. at APCTP Pohang, the visit of Y. J. at Perimeter Institute
for Theoretical Physics and the visit of I. K. at Simons Center for
Geometry and Physics.  We thank APCTP, PI and SCGP for hospitality.
This work has received support from the PHC Balaton program 30484YF,
from the European Programme IRSES UNIFY (Grant No 269217) and from the
People Programme (Marie Curie Actions) of the European Union's Seventh
Framework Programme FP7/2007-2013/ under REA Grant Agreement No
317089.

\appendix

   \section{The operator $\U $}
\la{AppendixA}

In this Appendix we collect some formulas about the action of the
operator $\U =U U_F$ which represents a finite super-conformal
transformation.  The operator is a product of a
$\mathfrak{su}(2,2)$-rotation in imaginary angle
\be \la{defU} U = e^{- \frac{\pi}{4} (P_0-K_0)} = e^{- \frac{\pi}{4}
(L^+_0-L^-_0)} =e^{ - \frac{\pi}{4} ( a_i^\dag b_i ^\dag + b_ia_i)} \;
\ee
and a unitary $\mathfrak{su}(4)$-rotation \be
\label{uwithfp}
U_F = e^{-{\pi\over 4} (R_{13} - R_{31} + R_{24}- R_{42})}=
e^{-\frac{\pi}{4} (c_i^\dagger d_{i}^\dagger- d_i c_{i})}\;.  \ee As
it was suggested in \cite{ADGN}, it is convenient to first to compute
the action of a rotation in an arbitrary angle $it$
  \be U_t = U_t ^\dag \equiv e^{ t ( a_i^\dag b_i ^\dag + b_ia_i)} .
  \ee
 The action of $U_t$ on the oscillators $a_i, a^\dag_i, b_i, b^\dag_i$
 is
   \be \la{thevolution} a_i(t ) \equiv U_t a _iU_t^{-1} = a_i \cos t -
   b_i^\dag \sin t , \quad b_i(t ) \equiv U_t b _iU_t^{-1} = b _i\cos
   t - a_i^\dag \sin t , \no \\
    a^\dag _i(t ) \equiv U_t a^\dag _iU_t^{-1} = a_i^\dag \cos t + b_i
    \sin t , \quad b^\dag _i(t ) \equiv U_t b^\dag _iU_t^{-1} =
    b_i^\dag \cos t + a _i \sin t .  \ee
     From here one easily obtains the normal form of the operator $U_t
     $ is \cite{ADGN}
  \be\la{Unormal} U_t \equiv e^{ t \( a^\dag b^\dag + ba\)}
   &=& {1\over \cos^2 t }\ e^{\tan t \, a^\dag b^\dag} (\cos t )^{ -\,
   a_i^\dag a- b^\dag_i b}\ e^{\tan t \, ba}, \ee
or, in terms of the Lie-algebra generators,
 \be U_t = e^{-t (L^+_0-L^-_0)}= {1\over \cos^2 t }\e^{-L^+_0\tan t}
 \cos(t)^{- 2E} e^{L_0^- \tan t}.  \ee
   Similarly one derives the normal form of the compact piece
   \re{uwithf} by introducing the rotation at angle $t$,
  \be\la{Unormalb} U_{t}^F
  \equiv e^{t \( c^\dag
  d^\dag + cd\)}
   &=& { \cos^2 t}\ e^{\tan t\, c^\dag d^\dag} (\cos t)^{ -\, c_i^\dag
   c_i- d^\dag_i d}\ e^{\tan t\, cd}.  \ee
In the normal form of the full operator, the $\cos t$ factors
     nicely cancel,
         \be\la{Unormaltot} \U_t & \equiv& e^{ t \( a^\dag b^\dag + ab + c^\dag
  d^\dag + cd \)}
  \no
  \\
  &=& e^{\tan t \, (a^\dag b^\dag+c^\dag d^\dag)} e^{ -
  \log \cos t (a^\dag a+ b^\dag b +c^\dag c+ d^\dag d} e^{\tan t \,
  (ab +cd)} .  \ee
  From \re{Unormaltot} one obtains the regularised expression for the
  conjugate vacuum $|\bar 0\> =|\bar 0\> _B\otimes |\bar 0\> _F$,
  \be
  |\bar 0\> \equiv
  \U^2 \ket&\approx & e^{ (a^\dag b^\dag + c^\dag d^\dag )/\e} |0\>
 \no
 \\
  & \approx & {e^{ a^\dag b^\dag /\e} \over \e^2} \ c^\dag_1 c^\dag_2\
  d^\dag_2 d^\dag_1\ket\, , \qquad \e\to 0.  \ee

  \section{Sending roots to infinity}

\label{apps:sep}

The limit $u\to\infty$ is delicate and can produce different results.
Here it is important that half of the roots are on shell and that we
send to infinity $k$ on-shell roots and $k$ off-shell roots.  We
proceed as follows: first send sequentially $k$ on-shell $\uu$-roots
to infinity so that the Bethe equations are satisfied in the process.
This is important, because otherwise the scalar product is not given
by a determinant.  Then we send $k$ off-shell $\vv$-roots to infinity.

Proceeding as in \cite{SL} (eq.  (3.24)) and taking into account that
$f(v_j) \approx e^{iG_\uu+i G_\vv- L/u} \approx e^{i(2N-L)/v_j}$ for
the $v$-roots, and as $f(u_k)\approx e^{ i G_\vv - i G_\uu} \approx
e^{ 0/u_k}$ because of the Bethe equations, one obtains the general
formula, when $K'= N-M'$ roots $\uu$ (on shell) and $K''=N-M''$ roots
$\vv$ (off shell) are sent to infinity:
 \be \la{BPSscpG} \lim_{v_{N-K+1}, \dots, v_N \to\infty}\[
 \left(\prod_{j=M''+1}^N v_j \right)\lim_{u_{N-K+1}, \dots, u_N
 \to\infty} \left( \prod_{j=M'+1}^N u_j \right)\langle
 \uu_N|\vv_N\rangle\]
       \no
 \\    =    (N-M')!(N-M'')!  \begin{pmatrix} L- M'-M''
\\
      N-M ''\end{pmatrix}
      \CA_{\uu_{M'}\cup \vv_{M''}}.
 \ee
where $ \CA_{\uu_{M'}\cup \vv_{M''}}$ is the determinant expression
giving the scalar product \cite{SL}.  Taking $K'=K''=k$ one obtains
the correct combinatorial factor from equation (\ref{magntriv})

\begin{eqnarray}
\label{magntrivv}
\langle\uu;k|k;\vv\rangle=(k!)^2 {L-2M \choose k} \langle\uu
|\vv\rangle\;.
\end{eqnarray}

  \section{The Spin Vertex as a Flipping Operator}
  \label{appendixflip}

In section we will justify the expression for the spin vertex \re{Aver} and explain why the expressions \re{eq1}, \re{eq2} give the correct  expression for the two- and three-point functions.

The propagators for the elementary fields have the following form:
\beq
\begin{split}
&\langle \bar S(y){S}(x)\rangle =  \frac{1}{ (x-y)^2}, \quad S = X, Y, Z,\\
&\langle \bar\Psi_{j b}(y){\Psi}_{ia}(x)\rangle =  i\delta_{ab}\sigma_{ij}^{\mu}\partial_{x^{\mu}}\frac{1}{ (x-y)^2}, \quad a,b=1,...,4, \quad i,j=1,2,\\
&\langle {\cal F}_{\rho \sigma}(y){\cal F}_{\mu \nu}(x) \rangle =
(\eta_{\nu\sigma}\partial_{\mu}\partial_{\rho}+\eta_{\mu\rho}\partial_{\nu}\partial_{\sigma}-\eta_{\mu\sigma}\partial_{\nu}\partial_{\rho}-
\eta_{\nu\rho}\partial_{\mu}\partial_{\sigma})\frac{1}{ (x-y)^2}.
\end{split}
\eeq
We have to show that the spin vertex formalism reproduce these propagators correctly, by means of the equation
\footnote{The ordering of the operators on the left hand side is chosen to ensure right sign for the fermionic propagator.}
\begin{align}
\label{eq11}
\langle {\cal O}_2(y){\cal O}_1(x) \rangle = \langle \CV_{12}|\,
e^{i( L^+_1 x+ L^+_2 y)} | \CO_2\rangle \otimes |{\cal O}_1\rangle \,.
\end{align}
First we establish the rule how the vertex transform the fields form the space $(2)$ to the space $(1)$. Using the representation of the elementary fields in terms of the oscillators
\beq
\begin{split}
&Z=|0\rangle, \quad
\bar{Z} = c_1^{\dagger}d_1^{\dagger}c_2^{\dagger}d_2^{\dagger}|0\rangle,\\
&Y=c_2^{\dagger}d_1^{\dagger}|0\rangle, \quad
\bar{Y} = c_1^{\dagger}d_2^{\dagger}|0\rangle,\\
&X=c_1^{\dagger}d_1^{\dagger}|0\rangle, \quad
\bar{X}=-c_2^{\dagger}d_2^{\dagger}|0\rangle,\\
&\Psi_{i1} = b_{i}^{\dagger}c_2^{\dagger}|0\rangle, \quad
\bar\Psi_{i1}=-a_{i}^{\dagger}c_1^{\dagger}d_2^{\dagger}d_1^{\dagger}|0\rangle,\\
&\Psi_{i2} =  -b_{i}^{\dagger}c_1^{\dagger}|0\rangle, \quad
\bar\Psi_{i 2}=-a_{i}^{\dagger}c_2^{\dagger}d_2^{\dagger}d_1^{\dagger}|0\rangle,\\
&\Psi_{i 3} =  b_{i}^{\dagger}c_1^{\dagger}c_2^{\dagger}d_1^{\dagger}|0\rangle, \quad
\bar\Psi_{i 3}=a_{i}^{\dagger}d_2^{\dagger}|0\rangle,\\
&\Psi_{i4} = b_{i}^{\dagger}c_1^{\dagger}c_2^{\dagger}d_2^{\dagger}|0\rangle, \quad
\bar\Psi_{i 4}=-a_{i}^{\dagger}d_1^{\dagger}|0\rangle,\\
&{F}_{ij} =-
b_{i}^{\dagger}b_{j}^{\dagger}c_1^{\dagger}c_2^{\dagger}|0\rangle, \quad
\bar{F}_{ij}= a_{i}^{\dagger}a_{j}^{\dagger}d_1^{\dagger}d_2^{\dagger}|0\rangle,
\end{split}
\eeq
we obtain by by direct computation
\beq
\begin{split}
&{}^{(2)}\langle S|U_F^2|V_{12}\rangle = |\bar{S}\rangle^{(1)},\quad
{}^{(2)}\langle \bar{S}|U_F^2|V_{12}\rangle = |S\rangle^{(1)}, \quad S = X,Y,Z\\
&{}^{(2)}\langle \Psi_{i a}|U_F^2|V_{12}\rangle = | \bar\Psi_{ia}\rangle^{(1)}, \quad
{}^{(2)}\langle \bar\Psi_{ia} |U_F^2|V_{12}\rangle = |\Psi_{i a}\rangle^{(1)},\\
&{}^{(2)}\langle F_{ij}|U_F^2|V_{12}\rangle = |\bar{F}_{ij}\rangle^{(1)}, \quad
{}^{(2)}\langle \bar{F}_{ij} |U_F^2|V_{12}\rangle = |F_{ij}\rangle^{(1)},\\
\end{split}
\eeq
where
\beq
\begin{split}
&U_F^2|V_{12}\rangle =e^{\sum\limits_{i=1,2}(b_i^{(1)\dagger}a_i^{(2)\dagger}-a_i^{(1)\dagger}b_i^{(2)\dagger}+
c_i^{(1)}c_i^{(2)\dagger}-d_i^{(1)}d_i^{(2)\dagger})}
c_1^{(1)\dagger}d_1^{(1)\dagger}c_2^{(1)\dagger}d_2^{(1)\dagger}|0\rangle^{(1)}|0\rangle^{(2)},
\end{split}
\eeq
and
\beq
\begin{split}
{\cal F}^{\mu\nu} = (\bar\sigma^{\mu\nu}\epsilon)_{ij} \bar{F}_{ij}-(\epsilon\sigma^{\mu\nu})_{ij}F_{ij},&
\quad i,j=1,2, \quad \mu,\nu=1,...,4,\\
\sigma^{\mu\nu} = \frac{1}{4}\Big(\sigma^{\mu}\bar\sigma^{\nu}-
\sigma^{\nu}\bar\sigma^{\mu}\Big),& \quad
\bar\sigma^{\mu\nu} = \frac{1}{4}\Big(\bar\sigma^{\mu}\sigma^{\nu} -
\bar\sigma^{\nu}\sigma^{\mu}\Big), \quad \epsilon_{12} = 1.
\end{split}
\eeq
This leads to the following expansion for the vertex
\beq
U^2_F|V_{12}\rangle = |\bar{S}^{(2)}_i\rangle|S_i^{(1)}\rangle +|S^{(2)}_i\rangle|\bar{S}^{(1)}_i\rangle +
|\bar\Psi^{(2)}_{ia}\rangle|\Psi^{(1)}_{i a}\rangle +
|\Psi^{(2)}_{ia}\rangle|\bar{\Psi}^{(1)}_{i a}\rangle +|\bar{F}^{(2)}_{ij}\rangle |F^{(1)}_{ij}\rangle
+|F^{(2)}_{ij}\rangle |\bar{F}^{(1)}_{ij}\rangle +...,
\eeq
where we assume summation over repeating indexes and three dots mean other possible states appearing in the vertex expansion, including those
not satisfying the zero central charge condition.

Now we are ready to compute the propagators using the \re{eq11}. We start with the scalars.
\beq
\begin{split}
&\langle \bar S(y){S}(x) \rangle = \langle \CV_{12}|
e^{i( L^+_{(1)} x+ L^+_{(2)} y)}|\bar S\rangle_{(2)} \otimes | {S}\rangle_{(1)}  =\langle {\cal V}_{12}|
e^{i( L^+_{(1)} x- L^+_{(1)} y)}|\bar S\rangle_{(2)} \otimes | {S}\rangle_{(1)} =\\
&\langle V_{12}|U_{F(1)}^2U_{(1)}^2
e^{i( L^+_{(1)} x- L^+_{(1)} y)}|\bar S\rangle_{(2)} \otimes | {S}\rangle_{(1)} =
\langle S|U^2 e^{i( L^+ x - L^+ y)}|S\rangle =\\ &\langle 0|U^2 e^{i( L^+ x - L^+ y)}|0\rangle= \frac{1}{(x-y)^2},
\end{split}
\eeq
where in order to get the last line we used  \re{corrtree}. For the fermions we'll consider one of the possible
propagators, the rest
can be computed absolutely analogously:
\beq
\begin{split}
&\langle \bar\Psi_{j 4}(y){\Psi}_{i4}(x) \rangle = -\langle {\cal V}_{12}|e^{i( L^+_{(1)} x+ L^+_{(2)} y)}a_{j}^{(2)\dagger}d_1^{(2)\dagger}|0\rangle_{(2)} b_{i}^{(1)\dagger} c_1^{(1)\dagger}c_2^{(1)\dagger}d_2^{(1)\dagger}|0\rangle_{(1)} =\\
&\langle 0|b_{i}d_2 c_2 c_1U^2 e^{i( L^+ x - L^+ y)}
b_{i}^{\dagger} c_1^{\dagger}c_2^{\dagger}d_2^{\dagger}|0\rangle =
 -\langle 0|U^2 e^{-i L^+ y} a_{j}^{\dagger}
b_{i}^{\dagger}e^{i L^+ x}|0\rangle =  \frac{i}{2}\partial_{\mu}\sigma^{\mu}_{ij}\langle 0|U^2 e^{i( L^+ x - L^+ y)}|0\rangle\\ &=\frac{i}{2}\partial_{\mu}\sigma^{\mu}_{ij}\frac{1}{(x-y)^2},
\end{split}
\eeq
where we used the explicit expression in terms of the oscillators for the $L^{+\mu} = -a_{i}^{\dagger}\bar\sigma_{ij}^{\mu}b_{j}^{\dagger}$ and the property of the $\sigma$ matrices
\beq
\sigma_{ij}^{\mu}(\bar\sigma_{\mu})_{kl} = -2\delta_{il}\delta_{jk}.
\eeq
Finally we compute the propagator for the strength field:
\beq
\begin{split}
&\langle F^{\rho\sigma}(y) F^{\mu\nu}(x)\rangle=\langle {\cal V}_{12}|e^{i L^+_{(1)} x}
e^{iL^+_{(2)}y}\Big((\bar\sigma^{\mu\nu}\epsilon)_{ij}a_{i}^{(2)\dagger}a_j^{(2)\dagger}d_1^{(2)\dagger}d_2^{(2)\dagger}+
(\sigma^{\mu\nu})_{ij}b_i^{(2)\dagger}b_{j}^{(2)\dagger}c_1^{(2)\dagger}c_2^{(2)\dagger}\Big)|0\rangle_{(2)}\otimes\\
&\Big((\bar\sigma^{\mu\nu}\epsilon)_{ij}a_{i}^{(1)\dagger}a_j^{(1)\dagger}d_1^{(1)\dagger}d_2^{(1)\dagger}+
(\sigma^{\mu\nu})_{ij}b_i^{(1)\dagger}b_{j}^{(1)\dagger}c_1^{(1)\dagger}c_2^{(1)\dagger}\Big)|0\rangle_{(1)}=\\
&-(\bar\sigma^{\mu\nu}\epsilon)_{ij}(\epsilon\sigma^{\rho\sigma})_{kl}
\langle 0|U^2 e^{-i L^+ y} a_{i}^{\dagger}a_{j}^{\dagger}b_{k}^{\dagger}b_{l}^{\dagger}e^{i L^+_{(1)} x}|0\rangle +
(\mu \leftrightarrow \rho,\nu \leftrightarrow \sigma)=\\
&\frac{1}{4}(\bar\sigma^{\mu\nu}\epsilon)_{ij}(\epsilon\sigma^{\rho\sigma})_{kl}
\sigma^{\kappa}_{ki}\sigma^{\omega}_{lj}\partial_{\kappa}\partial_{\omega}
\frac{1}{(x-y)^2}+(\mu \leftrightarrow \rho,\nu \leftrightarrow \sigma)=\\
&\frac{1}{8}(\bar\sigma^{\mu\nu}\epsilon)_{ij}(\epsilon\sigma^{\rho\sigma})_{kl}
\Big(\sigma^{\omega}_{lj}\sigma^{\kappa}_{ki}+\sigma^{\kappa}_{lj}\sigma^{\omega}_{ki}\Big)
\partial_{\kappa}\partial_{\omega}
\frac{1}{(x-y)^2}+(\mu \leftrightarrow \rho,\nu \leftrightarrow \sigma).\\
\end{split}
\eeq
Further we use the following identity:
\beq
\sigma_{ij}^{\mu}\sigma_{kl}^{\nu} + (\mu \leftrightarrow \nu) = -\eta^{\mu\nu}\bar\epsilon_{ik}
\bar\epsilon_{jl} + 4 \eta_{\kappa\omega}(\sigma^{\kappa\mu}\bar\epsilon)_{ik}(\bar\epsilon\bar\sigma^{\omega\nu})_{jl},
\eeq
where $\bar\epsilon_{12}=-1$.
It gives
\beq
\begin{split}
&\frac{1}{8}(\bar\sigma^{\mu\nu}\epsilon)_{ij}(\epsilon\sigma^{\rho\sigma})_{kl}
\Big(-\eta^{\kappa\omega}\bar\epsilon_{lk}
\bar\epsilon_{ji} + 4 \eta_{\tau\theta}(\sigma^{\tau\kappa}\bar\epsilon)_{lk}(\bar\epsilon\bar\sigma^{\theta\omega})_{ji}\Big)
\partial_{\kappa}\partial_{\omega}
\frac{1}{(x-y)^2}+(\mu \leftrightarrow \rho,\nu \leftrightarrow \sigma) =\\ &\Big(-\frac{\eta^{\kappa\omega}}{8}\Tr(\sigma^{\rho\sigma})\Tr(\bar\sigma^{\mu\nu})+\frac{\eta_{\tau\theta}}{2}
\Tr(\sigma^{\rho\sigma}\sigma^{\tau\kappa})\Tr(\bar\sigma^{\mu\nu}\bar\sigma^{\theta\omega})\Big)\partial_{\kappa}\partial_{\omega}
\frac{1}{(x-y)^2}+(\mu \leftrightarrow \rho,\nu \leftrightarrow \sigma).\\
\end{split}
\eeq
Next, noticing that $\Tr(\sigma^{\mu\nu})=\Tr(\bar\sigma^{\mu\nu})=0$ and also using the relations
\beq
\begin{split}
&\Tr(\sigma^{\mu\nu}\sigma^{\rho\sigma}) = -\frac{1}{2}\Big(\eta^{\mu\rho}\eta^{\nu\sigma}-\eta^{\mu\sigma}\eta^{\nu\rho}
+i\epsilon^{\mu\nu\rho\sigma}\Big),\\
&\Tr(\bar\sigma^{\mu\nu}\bar\sigma^{\rho\sigma}) = -\frac{1}{2}\Big(\eta^{\mu\rho}\eta^{\nu\sigma}-\eta^{\mu\sigma}\eta^{\nu\rho}
-i\epsilon^{\mu\nu\rho\sigma}\Big),
\end{split}
\eeq
we get
\beq
\begin{split}
&\langle F^{\rho\sigma}(y)F^{\mu\nu}(x)\rangle=\frac{\eta_{\tau\theta}}{8}\Big(\eta^{\rho\tau}\eta^{\sigma\kappa}-
\eta^{\rho\kappa}\eta^{\tau\sigma}+i\epsilon^{\rho\sigma\tau\kappa}\Big)\Big(\eta^{\mu\theta}\eta^{\nu\omega}-
\eta^{\mu\omega}\eta^{\nu\theta}-i\epsilon^{\mu\nu\theta\omega}\Big)\partial_{\kappa}\partial_{\omega}
\frac{1}{(x-y)^2}+\\&(\mu \leftrightarrow \rho,\nu \leftrightarrow \sigma)=\frac{1}{8}
\Big(\eta^{\sigma\kappa}\eta^{\mu\rho}\eta^{\nu\omega}-\eta^{\rho\nu}\eta^{\sigma\kappa}\eta^{\mu\omega}-
\eta^{\rho\kappa}\eta^{\mu\sigma}\eta^{\nu\omega}+\eta^{\rho\kappa}\eta^{\mu\omega}\eta^{\nu\sigma}
+i\epsilon^{\rho\sigma\mu\kappa}\eta^{\nu\omega}-i\epsilon^{\rho\sigma\nu\kappa}\eta^{\mu\omega}-\\
&i\epsilon^{\mu\nu\rho\omega}\eta^{\sigma\kappa}+i\epsilon^{\mu\nu\sigma\omega}\eta^{\rho\kappa}+
\eta_{\tau\theta}\epsilon^{\rho\sigma\tau\kappa}\epsilon^{\mu\nu\theta\omega}\Big)\partial_{\kappa}\partial_{\omega}
\frac{1}{(x-y)^2}+(\mu \leftrightarrow \rho,\nu \leftrightarrow \sigma).\\
\end{split}
\eeq
One can see that after taking into account symmetrization with respect to the permutation $(\mu \leftrightarrow \rho,\nu \leftrightarrow \sigma)$ and also $(\kappa \leftrightarrow \omega)$, all the terms proportional to $i$ cancel out. Decomposition of the Levi-Civita tensor contraction gives (we use convention $\epsilon^{0123}=1$)
\beq
\eta_{\tau\theta}\epsilon^{\rho\sigma\tau\kappa}\epsilon^{\mu\nu\theta\omega} = \eta^{\sigma\nu}\eta^{\rho\omega}\eta^{\kappa\mu}+
\eta^{\sigma\omega}\eta^{\rho\mu}\eta^{\kappa\nu}+\eta^{\rho\nu}\eta^{\sigma\mu}\eta^{\kappa\omega}-
\eta^{\sigma\omega}\eta^{\rho\nu}\eta^{\kappa\mu}-\eta^{\sigma\mu}\eta^{\rho\omega}\eta^{\kappa\nu}-\eta^{\rho\mu}\eta^{\sigma\nu}\eta^{\kappa\omega}.
\eeq
The terms proportional to $\eta^{\kappa\omega}$ cancel out due to equation of motion $\partial^2\frac{1}{(x-y)^2}=0$. Taking all this remarks into account we get final result:
\beq
\langle F^{\rho\sigma}(y)F^{\mu\nu}(x)\rangle=\frac{1}{2}\Big(\eta^{\sigma\kappa}\eta^{\mu\rho}\eta^{\nu\omega}-\eta^{\rho\nu}\eta^{\sigma\kappa}\eta^{\mu\omega}-
\eta^{\rho\kappa}\eta^{\mu\sigma}\eta^{\nu\omega}+\eta^{\rho\kappa}\eta^{\mu\omega}\eta^{\nu\sigma}\Big)\partial_{\kappa}\partial_{\omega}
\frac{1}{(x-y)^2}.
\eeq

The action of covariant derivatives in terms of oscillators is given by ${\cal D}_{ij}=a^{\dagger}_{i}b^{\dagger}_{j}$.
Thus, in case, when an elementary field belongs to the non-compact sector, the corresponding propagator can be obtained by taking appropriate number of derivatives contracted with right component of the sigma matrices, e.g.
\beq
\begin{split}
&\langle  \bar{Z}(y) {\cal D}_{ij}Z(x)\rangle = \langle \CV_{12}|
e^{i( L^+_{(1)} x+ L^+_{(2)} y)}|\bar Z\rangle_{(2)} \otimes |{\cal D}_{ij}Z\rangle_{(1)} \\ & = -\frac{i}{2}\sigma^{\mu}_{ji}\partial_{x^\mu}
\langle \CV_{12}|
e^{i( L^+_{(1)} x+ L^+_{(2)} y)}|\bar Z\rangle_{(2)} \otimes |Z\rangle_{(1)} = -\frac{i}{2}\sigma^{\mu}_{ji}\partial_{x^\mu}\frac{1}{(x-y)^2}.
\end{split}
\eeq

\providecommand{\href}[2]{#2}\begingroup\raggedright\endgroup

\end{document}